\documentclass[onecolumn,aps,floatfix,pra,superscriptaddress,nofootinbib]{revtex4}
\usepackage{color}
\usepackage{graphicx}
\usepackage{bm}
\usepackage{amsmath}
\usepackage{hyperref}
\usepackage{enumerate}
\usepackage{upgreek}
\usepackage[subnum]{cases}

\newcommand{\cl}{ \text{cl} }

\newcommand{\pa}{ \partial }
\newcommand{\hb}{ \hbar }
\newcommand{\si}{ \sigma }

\newcommand{\ga}{ \gamma }
\newcommand{\la}{ \langle }
\newcommand{\ra}{ \rangle }

\newcommand{\re}{ \text{Re} }
\newcommand{\sip}{ \text{sp} }
\newcommand{\erf}{ \text{erf} }
\newcommand{\MB}{ \text{MB} }
\newcommand{\BE}{ \text{BE} }
\newcommand{\FD}{ \text{FD} }
\begin{document}
\title{Dissipative two-identical-particle systems: diffraction and interference}
%
\author{S. V. Mousavi}
\email{vmousavi@qom.ac.ir}
\affiliation{Department of Physics, University of Qom, Ghadir Blvd., Qom 371614-6611, Iran}
\author{S. Miret-Art\'es}
\email{s.miret@iff.csic.es}
\affiliation{Instituto de F\'isica Fundamental, Consejo Superior de Investigaciones Cient\'ificas, Serrano 123, 28006 Madrid, Spain}
\begin{abstract}

Interference and diffraction of two-identical-particles are considered in the context of open quantum systems. This theoretical study is 
carried out within two approaches, the effective time-dependent Hamiltonian due to Caldirola-Kanai (CK) and the Caldeira-Leggett (CL) one where a 
master equation for the reduced density matrix is used under the presence of dissipation and temperature of the environment. 
Two simple but very illustrative examples are considered, diffraction by a single and two Gaussian slits by analyzing the mean square separation
between particles, single-particle probability density and the simultaneous detection probability or diffraction patterns.
Concerning the single Gaussian slit case, in the CK approach, the mean square separation drastically reduces with  friction, reaching a constant 
value due to the localization effect of friction. On the contrary, in the CL approach, temperature has an opposite effect to friction and this quantity increases.
Furthermore, there is a time-interval for which the joint detection probability 
is greater for fermions than for bosons. 
As has already been reported for non-dissipative systems, fermion bunching and boson anti-bunching are also observed. 
Decoherence process, loss of being indistinguishable, is settled gradually with time by increasing friction and temperature.
In the two Gaussian slits problem within the CK approach, the single-particle probability density behaves almost similarly for all kinds of particle-pairs displaying small overlapping between one-particle states. The differences among the three statistics decrease when dissipation increases.
However, in the opposite limit, fermions behave completely different from bosons which themselves behave like distinguishable particles.
This last behaviour is also seen when the interference pattern is considered by computing detection probability of both particles with two 
detectors, one fixed and the second mobile.

\end{abstract}

\maketitle

{\bf{Keywords}}: Identical particles, Dissipation, Caldirola-Kanai model, Caldeira-Leggett master equation, Diffraction, Interference, Joint detection probability, Mean square separation 


\section{Introduction}

In classical physics, identical particles are seen as mutually distinguishable; that is, any permutation bringing about an interchange of particles in
two different one-particle states is recognized to lead to a new, physically distinct state. Many times, we speak about Boltzmannian
particles obeying the Maxwell-Boltzmann (MB) statistics. In quantum theory, things are completely different since identical particles are not 
distinguishable. For such  indistinguishable particles, a factorization of the total wave function is not appropriate because 
an interchange  of any two particles leads to a wave function which has to be insensitive to such a permutation operation.  Now, depending on
the nature of identical particles, fermions and bosons, the total wave function is anti-symmetric and symmetric, respectively. As is well known, 
fermions follow the Fermi-Dirac (FD) statistics and bosons the Bose-Einstein (BE) statistics \cite{Baym-book-1969}. Furthermore, due to this symmetry, spatial correlations
exist even if the particles are supposedly noninteracting. This type of correlation is manifested through the so-called statistical inter-particle potential
which depends on the temperature through the thermal de Broglie wavelength \cite{Pathria-book-1988}.
For fermions and bosons, this potential is repulsive and attractive, respectively. There is also an intimate connections between these statistics 
and the intrinsic spin of those particles but we are not going to consider it here.

Interferometry and diffraction of matter waves is a very active field of research being a very valuable tool in examining the validity of 
quantum mechanics. Observed optical-like effects  are determined by the interaction 
between the corresponding particles and measuring devices. One of the  paradigmatic examples is the so-called Talbot effect,
that is, a near-field interference effect \cite{Salva-JCP-2007} due to the presence of two or more slits, or in general, any grating.  
In analogy to the so-called carpets reported in optics, quantum 
carpets \cite{Berry-PW-2001} are observed in the Fresnel region when gratings are illuminated by particle beams. 
These carpets are distorted  due to the specific particle-grating interaction 
(what has been termed the Talbot-Beeby effect \cite{Salva-JCP-2007}). When small clusters or big molecules are used, diffraction patterns
in the far field or Fraunhofer region are governed mainly by van der Waals interactions and strong reduction of the fringe visibility is 
observed \cite{Arndt-NatPhys-2007}. A way to reduce as much as possible such distortions is thanks to the well-known effect of quantum
reflection \cite{Wieland-Sci-2011, Salva-JPCL-2017, Salva-PRA-2018}. On the other hand, the effect of two-particle in this kind of 
studies has been extensively studied after the pioneer work by Hong et al. \cite{Hong-PRL-1987}; in particular, two photons. 
Identical massive particles' interference has been analyzed quite recently in several papers \cite{Bose-PRL-2002, Lim-NJP-2005, 
Sancho-EPJD-2014, MaGr-EPJD-2014}.
It has been argued that when one of the one-particle states has a zero, bunching and anti-bunching can occur for fermions and bosons,
respectively. \cite{MaGr-EPJD-2014} 
This unexpected effect can be seen by measuring detection probability of a couple of identical particles by two adjacent detectors: bosons 
(fermions) interfere locally destructively (constructively) and therefore anti-bunch (bunch). Furthermore, it has been shown that in the case 
of measurements by a single extended detector, there is no difference between the detection probabilities for different statistics meaning again 
bosons do not bunch and fermions do not anti-bunch \cite{MaGr-EPJD-2014}.
Concerning the role of quantum statistics in multi-particle decay dynamics, by studying the releasing of a pair of particles from a quantum trap, 
it has been concluded that the naive picture, in which identical bosons attract one another while identical fermions repel each other, does not work 
in predicting even the qualitative behavior of the pair \cite{MaGr-AP-2015}.

On the other hand, very few studies have  been devoted to analyze the effect of friction and temperature on the resulting interference and/or 
diffraction patterns. The main purpose of this kind of studies is to see how the decoherence process affects this open dynamics and, in this 
context, how robust is the symmetry properties of the wave function for indistinguishable particles.
Within the Caldirola-Kanai (CK) approach, where it takes into account friction without including environmental fluctuations, it has been shown 
in the two-slit problem for distinguishable particles how the friction leads to localization by using Bohmian trajectories \cite{Salva-AP-2014}.
This analysis is carried out in an analytical way for Gaussian slits because this approach keeps the linearity of the problem. Within the same 
context, entanglement indicators corresponding to pure states by using the nonlinear Schr\"odinger-Langevin wave equation 
\cite{Kostin, MoMi-AP-2018, MoMi-JPC-2018, MoMi-EPJP-2019, MoMi-Arxiv-2019} have been analyzed \cite{ZaPl-En-2018}. An alternative way
of dealing with the same issue and keeping the linearity is by means of the Caldeira-Leggett (CL) approach where a master equation for the reduced 
density in the coordinate representation and at high-temperatures is used \cite{Caldeira-PA-1983,Caldeira-book-2014}. 
Two simple but very illustrative examples, taking the one-particle states with considerable overlap, are considered, diffraction by a single and two Gaussian slits by analyzing the mean square separation
between distinguishable particles, bosons and fermions as well as the simultaneous detection probability or diffraction patterns. 
In the CK approach, the mean square separation drastically reduces with friction, reaching ultimately a constant value. 
On the contrary, in the CL approach, temperature has an opposite effect to friction and this quantity increases.
Furthermore, there is a time-interval for which the joint detection probability measured by an extended detector is greater for 
fermions than for bosons. As has already been reported for non-dissipative systems, fermion bunching and boson anti-bunching are also observed here for open two-particle systems. 
%
On the contrary, in the two Gaussian slit problem within  the CK approach, the interference pattern behave similarly for bosons and distinguishable 
particles whereas for fermions display a different behavior, reflecting the symmetry of the corresponding wave functions.
Recently, both approaches have been used for studying dissipative quantum backflow for distinguishable particles \cite{MoMi-arXiv-2019}.

In this work, our aim is to explore dissipative and thermal effects in the diffraction of indistinguishable particles (bosons and fermions) 
by one and two Gaussian slits. Due to the above-mentioned symmetry properties, three important quantities are going to be 
evaluated, the mean square separation (MSS) between particles, single-particle probability density and the simultaneous detection probability as analyzed in 
Refs. \cite{Sancho-EPJD-2014} and \cite{MaGr-EPJD-2014}. The decoherence process is ultimately studied  taken into account the mutual roles of friction and overlapping integral of one-particle states at a given time.
For this goal, 
this manuscript is organized as follows. In Section \ref{sec: tp_CK}, the two-particle CK equation from the corresponding 
one-particle equation is proposed. Section \ref{sec: tp_CL} is devoted to the evolution of the {\it pure} two-identical-particle state under 
the corresponding CL master equation. Section \ref{sec: D_sl} deals with diffraction of two-identical-particle state by a single Gaussian slit in 
both the CK and CL approaches. In section \ref{sec: tp-tl}, two-particle two-slit experiment will be studied within the CK approach.
Analytical and numerical results and discussion will be presented at the same time in the previous sections. In the last section, some 
concluding remarks are briefly listed. Finally, in an appendix, 
dissipative identical-particle systems are shown not to be properly described in the framework of the Schr\"{o}dinger-Langevin nonlinear 
wave equation because symmetry properties are no longer kept.

\section{The Caldirola-Kanai equation for two non-interacting particles} \label{sec: tp_CK}

The dissipative dynamics of two non-interacting particles with the same mass $m$ in the CK framework can be written as 
%
%
%
\begin{eqnarray} \label{eq: 2par_CK}
i \hb \frac{\pa }{\pa t} \Psi(x_1, x_2; t)  &=& \bigg[ - e^{-2\ga t} \frac{\hb^2}{2m} 
\left( \frac{\pa^2}{\pa x_1^2} + \frac{\pa^2}{\pa x_2^2}\right) + e^{2\ga t} ( V(x_1) + V(x_2) )
  \bigg] \Psi(x_1, x_2; t)
\end{eqnarray}
where for identical particles the wave function $ \Psi(x_1, x_2; t) $ must have a given symmetry; it must be symmetric (anti-symmetric) 
under the exchange of identical bosons (fermions) obeying respectively the BE and FD statistics 
\cite{Baym-book-1969}. 
Here $ \ga = \eta/ 2m $ is the relaxation constant \cite{Caldeira-PA-1983} defined versus the damping constant $ \eta $.
If the initial wave function is expressed as  
\begin{eqnarray} \label{eq: 2p-psi0}
\Psi_{\pm}(x_1, x_2, 0) &=& N_{\pm} ( \psi_0(x_1) \phi_0(x_2) \pm \phi_0(x_1) \psi_0(x_2) )
\end{eqnarray}
where $\psi$ and $\phi$ are one-particle states and sub-indices $+$ and $-$ stand respectively for bosons and fermions, then due to the linearity 
of the wave equation (\ref{eq: 2par_CK}), the symmetric and anti-symmetric solutions can be written as
\begin{eqnarray} \label{eq: 2p-psit}
\Psi_{\pm}(x_1, x_2, t) &=& N_{\pm} ( \psi(x_1, t) \phi(x_2, t) \pm \phi(x_1, t) \psi(x_2, t) )
\end{eqnarray}
at later times, where $\psi(x, t)$ and $\phi(x, t)$ fulfill the corresponding  one-particle CK wave equation. Apart from a phase factor, 
the normalization constants $ N_{\pm} $ are given by
\begin{eqnarray} \label{eq: normalization}
N_{\pm} &=& \frac{1}{ \sqrt{ 2(1 \pm | \la \psi_0 | \phi_0 \ra |^2)  } }
\end{eqnarray}
where we have assumed that the one-particle wave functions $\psi$ and $\phi$ are normalized to unity as well as used the fact that the 
interference term $ \la \psi(t) | \phi(t) \ra $ is independent on time. This can be easily deduced from the one-particle CK wave equation and 
square-integrability of the one-particle wave functions,
\begin{eqnarray} \label{eq: interference}
\frac{d}{d t} \la \psi(t) | \phi(t) \ra &=& \int_{-\infty}^{\infty} dx ~ \frac{\pa}{\pa t} [ \psi^*(x, t) \phi(x, t) ]
 =
e^{-2\ga t} \frac{i \hb}{2m} \int_{-\infty}^{\infty} dx \left\{ - \frac{\pa^2 \psi^*}{\pa x^2} \phi + \psi^* \frac{\pa^2 \phi}{\pa x^2}  \right\} = 0  .
\end{eqnarray}

In the following we are going to provide  expressions for two important quantities in this context, namely, the mean square separation (MSS) 
between particles and simultaneous detection probability (i) by a single non-ideal detector located at the origin and (ii) by two detectors located symmetrically around the origin. The widths of all detectors are 
the same, $2d$. This means we provide an expression for the probability of finding both particles simultaneously in the range $[-d, d]$ around the origin in the first case; and the same quantity for finding particles in two distinct regions of the same width $2d$ located symmetrically around the origin in the second case.
Finally, by tracing out over the coordinate of one of the particles, some expressions for single-particle probability density and its corresponding
probability current density fulfilling a continuity equation are analyzed.

\subsection{Mean square separation between particles}

One of the fundamental quantities in this context is the expectation value of square distance between particles, 
$ \la \Psi| ( \hat{x}_1 - \hat{x}_2 )^2 | \Psi \ra$, for different statistics, which reads as
\begin{eqnarray} \label{eq: mss_CK}
\la ( \hat{x}_1 - \hat{x}_2 )^2 \ra_{\pm} &=& 
\la \Psi_{\pm}| ( \hat{x}_1 - \hat{x}_2 )^2 | \Psi_{\pm} \ra
= 2 |N_{\pm}|^2 \bigg( \la x^2 \ra_{\psi} + \la x^2 \ra_{\phi} - 2 \la x \ra_{\psi} \la x \ra_{\phi} 
\mp 2 | \la x \ra_{\psi \phi} |^2
\pm 2 \text{Re} \{ \la x^2 \ra_{\psi\phi} \la \phi|\psi\ra \} \bigg) 
\nonumber \\
& \equiv &
2 |N_{\pm}|^2 \bigg( \la ( \hat{x}_1 - \hat{x}_2 )^2 \ra_{\MB} \mp 2 | \la x \ra_{\psi \phi} |^2 \pm 2 \text{Re} \{ \la x^2 \ra_{\psi \phi} \la \phi|\psi\ra \} \bigg)
\end{eqnarray}
where, $ \la ( \hat{x}_1 - \hat{x}_2 )^2 \ra_{\MB} \equiv \la \psi \phi |( \hat{x}_1 - \hat{x}_2 )^2 | \psi \phi \ra $ stands for the expectation value 
of the square separation between the two distinguishable particles obeying the MB statistics; analogously, $ \la \cdots \ra_{\psi} 
\equiv \la \psi | \cdots | \psi \ra $ and $ \la \cdots \ra_{\psi \phi} \equiv \la \psi | \cdots | \phi \ra $.
%
%

\subsection{Simultaneous detection probability}

Consider now a single detector located at the origin with a width $2d$. Then, the {\it ratio} of simultaneous detection probability of indistinguishable particles to the distinguishable ones is given by \cite{MaGr-EPJD-2014}
\begin{eqnarray}  
p_{\pm}(t) &=& 
\frac{ p_{\substack{\BE \\ \FD}}(t)
}{ p_{\MB}(t) }= 
\frac{ \int_{-d}^{d} dx_1 \int_{-d}^{d} dx_2 | \Psi_{\pm}(x_1, x_2, t) |^2  }{ \int_{-d}^{d} dx_1 \int_{-d}^{d} dx_2 ~~\frac{1}{2} ( |\psi(x_1, t)|^2 |\phi(x_2, t)|^2 + |\psi(x_2, t)|^2 |\phi(x_1, t)|^2 ) } \\
&=&
2 N_{\pm}^2 
\left\{  
1 \pm \frac{ \left| \int_{-d}^{d} dx ~ \psi^*(x, t)\phi(x, t) \right|^2  }
{ \int_{-d}^{d} dx |\psi(x, t)|^2 \int_{-d}^{d} dx |\phi(x, t)|^2 }
\right\} \label{eq: detprob_CK}
\end{eqnarray}
where, in the second line, Eq. (\ref{eq: 2p-psit}) has been used. 
Note that for distinguishable particles obeying the MB statistics, the corresponding probability density is expressed as 
\begin{eqnarray} \label{eq: rhoMB_CK}
p_{\MB}(t) = | \Psi_{\MB}(x_1, x_2, t) |^2 &=& \frac{1}{2} ( |\psi(x_1, t)|^2 |\phi(x_2, t)|^2 + |\phi(x_1, t)|^2 |\psi(x_2, t)|^2 )
\end{eqnarray}
Just for completeness, we mention that if instead of a single detector, one considers two detectors  with the same width $2d$ located 
symmetrically around the origin at positions $D$ and $-D$ respectively, then the relative simultaneous detection probability for finding a 
particle by the first detector  and a particle by the second detector  is given by 
\begin{eqnarray}  
p'_{\pm}(t) &=& 
\frac{ p'_{\substack{\BE \\ \FD}}(t)
}{ p'_{\MB}(t) }= 
\frac{ \int_{D-d}^{D+d} dx_1 \int_{-D-d}^{-D+d} dx_2 | \Psi_{\pm}(x_1, x_2, t) |^2  }{ \int_{D-d}^{D+d} dx_1 \int_{-D-d}^{-D+d} dx_2 ~~\frac{1}{2} ( |\psi(x_1, t)|^2 |\phi(x_2, t)|^2 + |\psi(x_2, t)|^2 |\phi(x_1, t)|^2 ) } 
\\ \nonumber
\\ 
&=&
2 N_{\pm}^2 
\left\{  
1 \pm 2 
\frac{ \text{Re} \left\{\int_{D-d}^{D+d} dx_1 ~ \psi^*(x_1, t) \phi(x_1, t) \int_{-D-d}^{-D+d} dx_2 ~ \phi^*(x_2, t) \psi(x_2, t) \right\}  }
{ \int_{D-d}^{D+d} dx_1 |\psi(x_1, t)|^2 \int_{-D-d}^{-D+d} dx_2 |\phi(x_2, t)|^2 + \int_{D-d}^{D+d} dx_1 |\phi(x_1, t)|^2 \int_{-D-d}^{-D+d} dx_2 |\psi(x_2, t)|^2 }
\right\} \label{eq: detprob_CK_prime}
\end{eqnarray}
where we have preferred to use $ p'_{\pm}(t) $ to distinguish it from the previous case where one detector is only considered. 
Note that for $D=0$ this result reduces to the former result given by Eq. (\ref{eq: detprob_CK}). 
When the widths of detectors are negligible in comparation to their distance, $d \ll D$, which corresponds to two point detectors, 
then Eq. (\ref{eq: detprob_CK_prime}) reduces to
\begin{eqnarray}  \label{eq: detprob_CK_prime-point}
p'_{\pm}(t) 
&=&
2 N_{\pm}^2 \left[  
1 \pm 2 
\text{Re} 
\left\{
\left(  
\frac{ \psi(D, t) ~ \phi(-D, t) }{ \psi(-D, t) ~ \phi(D, t) } + 
\frac{\psi^*(-D, t) ~ \phi^*(D, t) }{ \psi^*(D, t) ~ \phi^*(-D, t) }
 \right)^{-1}
\right\} 
\right]
\end{eqnarray}
In the following, numerical results for the two detector scheme are provided.

\subsection{The continuity equation for reduced (single-particle) densities}

From the two-particle CK equation (\ref{eq: 2par_CK}), one can easily obtain the continuity equation written as
\begin{eqnarray} \label{eq: con_eq}
\frac{\pa}{\pa t} |\Psi|^2 + \frac{\hb}{m} e^{-2\ga t} \sum_k \frac{\pa}{\pa x_k} \text{Im} 
\left\{ \Psi^* \frac{\pa \Psi}{\pa x_k}  \right\} &=& 0  .
\end{eqnarray}
By integrating this equation over $x_2$, we have that
\begin{eqnarray} 
\frac{\pa}{\pa t} \int dx_2 ~ |\Psi(x, x_2, t)|^2 + \frac{\hb}{m} e^{-2\ga t} \frac{\pa}{\pa x}
\int dx_2 ~ \text{Im} 
\left\{ \Psi^*(x, x_2, t) \frac{\pa \Psi(x, x_2, t)}{\pa x}  \right\} &=& 0
\end{eqnarray}
which from Eq. (\ref{eq: 2p-psit}), the continuity equation for the reduced density can be written as
\begin{eqnarray} \label{eq: con_eq_1p}
\frac{\pa \rho_{\sip}(x, t)}{\pa t} + \frac{\pa j_{\sip}(x, t)}{\pa x} &=& 0
\end{eqnarray}
with
\begin{numcases}~
\rho_{\sip}(x, t) = |N_{\pm}|^2 ( |\psi(x, t)|^2 + |\phi(x, t)|^2 \pm 2 \text{Re}[ \la \psi | \phi \ra \phi^*(x, t) \psi(x, t) ]  ) \label{eq: rhosp}
\\
j_{\sip}(x, t) = |N_{\pm}|^2\frac{\hb}{m} e^{-2\ga t} \text{Im} \left\{ \psi^* \frac{\pa \psi}{\pa x} + \phi^* \frac{\pa \phi}{\pa x} \pm \la \phi|\psi \ra \psi^* \frac{\pa \phi}{\pa x}
\pm \la \psi|\phi \ra \phi^* \frac{\pa \psi}{\pa x}
\right\}
\end{numcases}
being the single-particle (sp) probability density and probability current density, respectively. Interference effects are noticeable in both expressions.

\section{The Caldeira-Leggett equation for two non-interacting particles} \label{sec: tp_CL}

So far we have only taken into account dissipative aspects of the environment through the well-known CK formalism. Thermal fluctuations due to the 
environment can also be considered following the CL formalism \cite{Caldeira-PA-1983, Caldeira-book-2014}. 
In this framework, the master equation describing the evolution of the reduced density matrix $ \rho $ in the coordinate representation and 
at high-temperatures reads 
\begin{eqnarray} \label{eq: CL eq}
\frac{\pa \rho}{\pa t} &=& \left[ - \frac{\hb}{2mi} \left( \frac{\pa^2}{\pa x^2} - \frac{\pa^2}{\pa x'^2} \right) - \ga (x-x') \left( \frac{\pa}{\pa x} - \frac{\pa}{\pa x'} \right)
+ \frac{ V(x) - V(x') }{ i\hb }
- \frac{D}{\hb^2} (x-x')^2 \right] \rho(x, x', t) 
\\
& \equiv & \mathcal{L}(x, x') \rho(x, x', t) 
\end{eqnarray}
with $ D = 2 m \ga k_B T $ being the diffusion coefficient; $k_B$ is the Boltzman factor and $T$ the temperature of the environment. 
If $\rho_1(x, x', t)$ and $\rho_2(x, x', t)$ are two one-particle states, due to the linearity of the operator $ \mathcal{L}(x, x') $ appearing 
in the one-particle CL equation (\ref{eq: CL eq}), the corresponding  master equation is then written as
\begin{eqnarray} \label{eq: 2p-CL eq}
\frac{\pa}{\pa t} [ \rho_1(x_1, x_1', t) \rho_2(x_2, x_2', t) ] &=& 
[ \mathcal{L}(x_1, x_1') + \mathcal{L}(x_2, x_2')  ] [ \rho_1(x_1, x_1', t) \rho_2(x_2, x_2', t) ]
\end{eqnarray}
for the evolution of the product state  $ \rho_1(x_1, x_1', t) \rho_2(x_2, x_2', t) $.

Consider now a system of two identical particles described by the initial density matrix
\begin{eqnarray} 
\rho_{\pm}(x_1, x_1'; x_2, x_2', 0) &=& \la x_1, x_2| \Psi_{\pm}(0) \ra \la \Psi_{\pm}(0) | x_1', x_2' \ra 
\\
&=& N_{\pm}^2 [ \psi_0(x_1) \phi_0(x_2) \pm \phi_0(x_1) \psi_0(x_2) ] [ \psi_0^*(x_1') \phi_0^*(x_2') \pm \phi_0^*(x_1') \psi_0^*(x_2') ]
\nonumber \\
&\equiv& N_{\pm}^2 [ \rho_{aa}(x_1, x_1', 0) \rho_{bb}(x_2, x_2', 0) \pm \rho_{ab}(x_1, x_1', 0) \rho_{ba}(x_2, x_2', 0)
\nonumber \\
&& \quad~ 
\pm \rho_{ba}(x_1, x_1', 0) \rho_{ab}(x_2, x_2', 0) + \rho_{bb}(x_1, x_1', 0) \rho_{aa}(x_2, x_2', 0) ]
\label{eq: rho0}
\end{eqnarray}
corresponding to the {\it pure} state given by Eq. (\ref{eq: 2p-psi0}). 
In the third line we have used the notation
\begin{numcases}~
\rho_{aa}(x, x', 0) = \psi_0(x) \psi_0^*(x')
\\
\rho_{ab}(x, x', 0) = \psi_0(x) \phi_0^*(x')
\\
\rho_{ba}(x, x', 0) = \phi_0(x) \psi_0^*(x')
\\
\rho_{bb}(x, x', 0) = \phi_0(x) \phi_0^*(x')   .
\end{numcases}
Each term of Eq. (\ref{eq: rho0}) corresponds to a product state evolving in time according to Eq. (\ref{eq: 2p-CL eq}). 
As a consequence, the evolution of each one-particle state like $ \rho_{ab}(x, x', 0) $ is given by the one-particle CL equation (\ref{eq: CL eq}). 
For Gaussian one-particle states, this equation can be solved by doing a coordinates transformation $ (x, x') \rightarrow (r, R) $, with $r=x-x'$ and 
$ R=(x+x')/2 $, taking a partial Fourier transform with respect to the coordinate $R$, solving the resulting equation and finally taking the 
inverse Fourier transform to obtain the density matrix in the coordinate representation \cite{Ve-PRA-1994&VeKuGh-PA-1995}. 
Following this procedure, quantum dissipative backflow for the superposition of two Gaussian wave packets has been studied in 
the CL framework \cite{MoMi-arXiv-2019}. 

Note that for distinguishable particles obeying the MB statistics, the density matrix in the configuration space is given by
\begin{eqnarray} \label{eq: rhoMB_CL}
\rho_{\MB}(x_1, x_1'; x_2, x_2', t) &=& \frac{1}{2} [ \rho_{aa}(x_1, x_1', t) \rho_{bb}(x_2, x_2', t) + \rho_{bb}(x_1, x_1', t) \rho_{aa}(x_2, x_2', t) ]
\end{eqnarray}
where $ \rho_{aa}(x_1, x_1', t) $ is the time evolution of the state $ \rho_{aa}(x_1, x_1', 0) $ under Eq. (\ref{eq: CL eq}).

Since our aim is the computation of detection probability and mean square separation, diagonal elements of one-particle density matrices
are only needed.

\subsection{Mean square separation between particles}

As before, for the MSS we have
\begin{eqnarray} 
\la ( \hat{x}_1 - \hat{x}_2 )^2 \ra_{\pm} &=& \text{Tr}[ ( \hat{x}_1 - \hat{x}_2 )^2 \hat{\rho}_{\pm}(t) ]
= \int_{-\infty}^{\infty} dx_1 \int_{-\infty}^{\infty} dx_2 ~ \la x_1, x_2 | ( \hat{x}_1 - \hat{x}_2 )^2 \hat{\rho}_{\pm}(t) | x_1, x_2 \ra 
\\
& = &
2 |N_{\pm}|^2 \bigg( \la ( \hat{x}_1 - \hat{x}_2 )^2 \ra_{\MB} \mp 2 | \la x \ra_{ab} |^2 \pm 2 \text{Re} \left\{ \la x^2 \ra_{ab} \int_{-\infty}^{\infty} dx ~ \rho_{ba}(x, x, t) \right\} \bigg) \label{eq: mss_CL}
\end{eqnarray}
where
\begin{numcases}~
\la ( \hat{x}_1 - \hat{x}_2 )^2 \ra_{\MB} = \la x^2 \ra_{aa} + \la x^2 \ra_{bb} - 2 \la x \ra_{aa} \la x \ra_{bb} \label{eq: mss_CL_MB}
\\
\la \cdots \ra_{ab} = \int_{-\infty}^{\infty} dx ~ ( \cdots ) ~ \rho_{ab}(x, x, t)
\end{numcases}

\subsection{Simultaneous detection probability}

Again, the ratio of simultaneous detection probability of indistinguishable particles to the distinguishable ones is computed in the CL framework 
to give 
\begin{eqnarray} 
p_{\pm}(t) &=& 
\frac{ p_{\substack{\BE \\ \FD}}(t)
}{ p_{\MB}(t) }= 
\frac{ \int_{-d}^{d} dx_1 \int_{-d}^{d} dx_2 ~ \rho_{\pm}(x_1, x_1; x_2, x_2, t)  }{ \int_{-d}^{d} dx_1 \int_{-d}^{d} dx_2 ~ \rho_{\MB}(x_1, x_1; x_2, x_2, t) } \\
&=&
2 N_{\pm}^2 
\left\{  
1 \pm \frac{ \left| \int_{-d}^{d} dx ~ \rho_{ab}(x, x, t) \right|^2  }
{ \int_{-d}^{d} dx ~ \rho_{aa}(x, x, t) \int_{-d}^{d} dx ~ \rho_{bb}(x, x, t) }
\right\} \label{eq: detprob_CL}
\end{eqnarray}
where the detector is located at the origin with a width of  $2d$.

\section{Results and discussion}

In this Section, diffraction of a two-identical-particle system by a single and two Gaussian slits is analyzed. In the following, numerical calculations 
are carried out in a system of units where $ m = \hb = 1$.

\subsection{Diffraction by a single Gaussian slit} \label{sec: D_sl}

The assumption of the Gaussian slit is due to Feynman \cite{Feynman-book-1965} which converts the problem to an analytical one. Otherwise, 
the corresponding analysis calls for the numerical integration of Fresnel functions \cite{Sa-JPB-2010}.

\subsubsection{The Caldirola-Kanai approach}

The initial one-particle wave packets $\psi$ and $\phi$ as two co-centred Gaussian wave packets with the same center $x_0$, kick momenta
$p_0$ and $\bar{p}_0$ and widths $\si_0$ and $\bar{\si}_0$ are assumed to be 
\begin{numcases}~
\psi_0(x) = \frac{1}{(2\pi \si_0^2)^{1/4}} \exp \left[ - \frac{(x-x_0)^2}{4\si_0^2} + i \frac{p_0}{\hb} (x-x_0) \right] \label{eq: psi0}
\\
\phi_0(x) = \frac{1}{(2\pi \bar{\si}_0^2)^{1/4}} \exp \left[ - \frac{(x-x_0)^2}{4\bar{\si}_0^2} + i \frac{\bar{p}_0}{\hb} (x-x_0) \right] \label{eq: phi0}   .
\end{numcases}
Then, the overlap integral is given by
\begin{eqnarray} \label{eq: ovint}
\la \phi_0| \psi_0 \ra &=& \sqrt{ \frac{2 \si_0 \bar{\si}_0 }{\si_0^2 + \bar{\si}_0^2 } } ~
 \exp \left[ - \frac{ \si_0^2 \bar{\si}_0^2 }{\si_0^2 + \bar{\si}_0^2} \frac{(p_0-\bar{p}_0)^2}{\hb^2} \right]
\end{eqnarray}
%
%
%
and the solution of the corresponding one-particle CK equation  
\begin{eqnarray} \label{eq: 1par_CK}
i \hb \frac{\pa }{\pa t} \psi(x, t)  &=& \bigg[ - e^{-2\ga t} \frac{\hb^2}{2m} \frac{\pa^2}{\pa x^2}  + e^{2\ga t} V(x) \bigg] \psi(x, t)
\end{eqnarray}
for the free propagation of the initial Gaussian wave packet (\ref{eq: psi0}) reads  as \cite{MoMi-JPC-2018}
\begin{eqnarray} \label{eq: spwf_Gauss}
\psi(x, t) &=& \frac{1}{(2\pi s_t^2)^{1/4}} \exp \left[ - \frac{(x-x_t)^2}{4\si_0 s_t} + i \frac{p_0}{\hb} (x-x_t) + \frac{i}{\hb} \mathcal{A}_{\cl}(t) \right]
\end{eqnarray}
where $s_t$, $x_t$ and $\mathcal{A}_{\cl}(t)$ are respectively the complex width, classical trajectory of the center of the wave packet and 
classical action given respectively by
\begin{numcases}~
s_t = \sigma_0 \left( 1 + i \frac{ \hbar}{2m\sigma_0^2} \uptau(t) \right) \label{eq: st},  \\
x_t = x_0 + \frac{p_0}{m} \uptau(t) ,  \\
\mathcal{A}_{\cl}(t) = \frac{p_0^2}{2m} \uptau(t) ,
\end{numcases}
with
\begin{eqnarray} \label{eq: uptau}
\uptau(t) &=& \frac{1-e^{-2\ga t}}{2\ga}    .
\end{eqnarray}
The complex width and center of the wave packet $\phi(x, t)$ are respectively denoted by $\bar{s}_t$ and $\bar{x}_t$. 


Now from the previous analysis one obtains that
\begin{numcases}~
\la x \ra_{\psi \psi} = x_t , \\
\la x \ra_{\phi \phi} = \bar{x}_t , \\
\la x^2 \ra_{\psi \psi} = \si_t^2 + x_t^2 , \\
\la x^2 \ra_{\phi \phi} = \bar{\si}_t^2 + \bar{x}_t^2 , \\
\la x \ra_{\psi \phi} =  \frac{\beta}{2\theta \sqrt{2 \theta s_t^* \bar{s}_t} } \exp\left[ \alpha + \frac{\beta^2}{4\theta}  \right],  \\
\la x^2 \ra_{\psi \phi} = \frac{\beta^2 + 2\theta}{4\theta^2 \sqrt{2 \theta s_t^* \bar{s}_t} } \exp\left[ \alpha + \frac{\beta^2}{4\theta}  \right] , 
\end{numcases}
where
\begin{numcases}~
\si_t = \si_0 \sqrt{ 1 + \frac{\hb^2}{4m^2\si_0^4} \uptau(t)^2 } ~, \label{eq: sigmat} \\
\bar{\si}_t = \bar{\si}_0 \sqrt{ 1 + \frac{\hb^2}{4m^2\bar{\si}_0^4} \uptau(t)^2 } ~, \label{eq: delt}
\end{numcases}
are the time dependent widths of the wave packets $\psi(x, t)$ and $\phi(x, t)$ respectively with
\begin{numcases}~
\alpha = - \frac{i}{ \hb } \Delta \mathcal{A}_{\cl}(t) + \frac{i}{ \hb } ( p_0 x_t - \bar{p}_0 \bar{x}_t )
- \frac{x_t^2}{4\si_0 s_t^*} - \frac{\bar{x}_t^2}{4\bar{\si}_0 \bar{s}_t} ,  \\
\beta = -\frac{i(p_0-\bar{p}_0)}{ \hb } + \frac{x_t}{2\si_0 s_t^*} + \frac{\bar{x}_t}{2\bar{\si}_0 \bar{s}_t} ,  \\
\theta = \frac{1}{4\si_0 s_t^*} + \frac{1}{4\bar{\si}_0 \bar{s}_t}   ,
\end{numcases}
where $\Delta \mathcal{A}_{\cl}(t)$ is the difference between the classical actions for the component wavepackets $\psi$ and $\phi$.
Thus, by using these relations, the mean square separation is computed through Eq.  (\ref{eq: mss_CK}).


On the other hand, for the one-particle detection probability one obtains that
\begin{eqnarray}
\int_{-d}^{d} dx |\psi(x, t)|^2 = 
\frac{1}{2} \left\{ \erf\left[ \frac{x_t +d}{\sqrt{2} \si_t}  \right] - \erf\left[ \frac{x_t - d}{\sqrt{2} \si_t}  \right] \right\}
\end{eqnarray}
and for the overlap integral in the detector region $ [-d, d] $ 
\begin{eqnarray}
\int_{-d}^{d} dx ~ \psi^*(x, t)\phi(x, t) &=& 
\sqrt{ \frac{\si_0 \bar{\si}_0}{ \si_0 s_t^* + \bar{\si}_0 \bar{s}_t } } 
~ e^{b_1(t)} ~ \{ \erf[b_2(t)] - \erf[b_3(t)] \}
\end{eqnarray}
where $\erf(\cdots)$ is the error function and 
\begin{numcases}~
b_1(t) =
\frac{
4i\hb [ ( \si_0 s_t^* + \bar{\si}_0 \bar{s}_t ) \Delta \mathcal{A}_{\cl}(t) - ( \si_0 s_t^* p_0 + \bar{\si}_0 \bar{s}_t \bar{p}_0 ) \Delta x_t ] + \hb^2 \Delta x_t^2 + 4 \si_0 \bar{\si}_0 s_t^* \bar{s}_t (p_0-\bar{p}_0)^2}
{ 4\hb^2( \si_0 s_t^* + \bar{\si}_0 \bar{s}_t ) } 
\\
b_2(t) = 
\frac{
-\hb [ \si_0 s_t^* (\bar{x}_t - d) + \bar{\si}_0 \bar{s}_t (x_t-d) ] + 2 i \si_0 \bar{\si}_0 s_t^* \bar{s}_t (p_0 - \bar{p}_0)}
{ 2\hb \sqrt{ \si_0 \bar{\si}_0 s_t^* \bar{s}_t ( \si_0 s_t^* + \bar{\si}_0 \bar{s}_t ) }  }
\\
b_3(t) =
\frac{
-\hb [ \si_0 s_t^* (\bar{x}_t + d) + \bar{\si}_0 \bar{s}_t (x_t+d) ] + 2 i \si_0 \bar{\si}_0 s_t^* \bar{s}_t (p_0 - \bar{p}_0)}
{ 2\hb \sqrt{ \si_0 \bar{\si}_0 s_t^* \bar{s}_t ( \si_0 s_t^* + \bar{\si}_0 \bar{s}_t ) }  }
\end{numcases}
$ \Delta x_t $ being the distance between centres of the component wavepackets $\psi$ and $\phi$ i.e., $ \Delta x_t = x_t - \bar{x}_t $.
By using these relations in Eq. (\ref{eq: detprob_CK}), the detection probability is easily reached in the CK framework.

In order to carry out numerical calculations, the following initial conditions are chosen: $ \si_0 = 1 $, $ x_0 = 0 $, $ p_0 = 3 $, $ \bar{\si}_0 = 0.9 $ and $\bar{p}_0 = p_0 $.
These conditions mean that the wave packets $\psi$ and $\phi$ have considerable overlap. Thus, one expects the behaviour of indistinguishable 
and distinguishable particles become completely different. 
\begin{figure} 
	\centering
	\includegraphics[width=12cm,angle=-0]{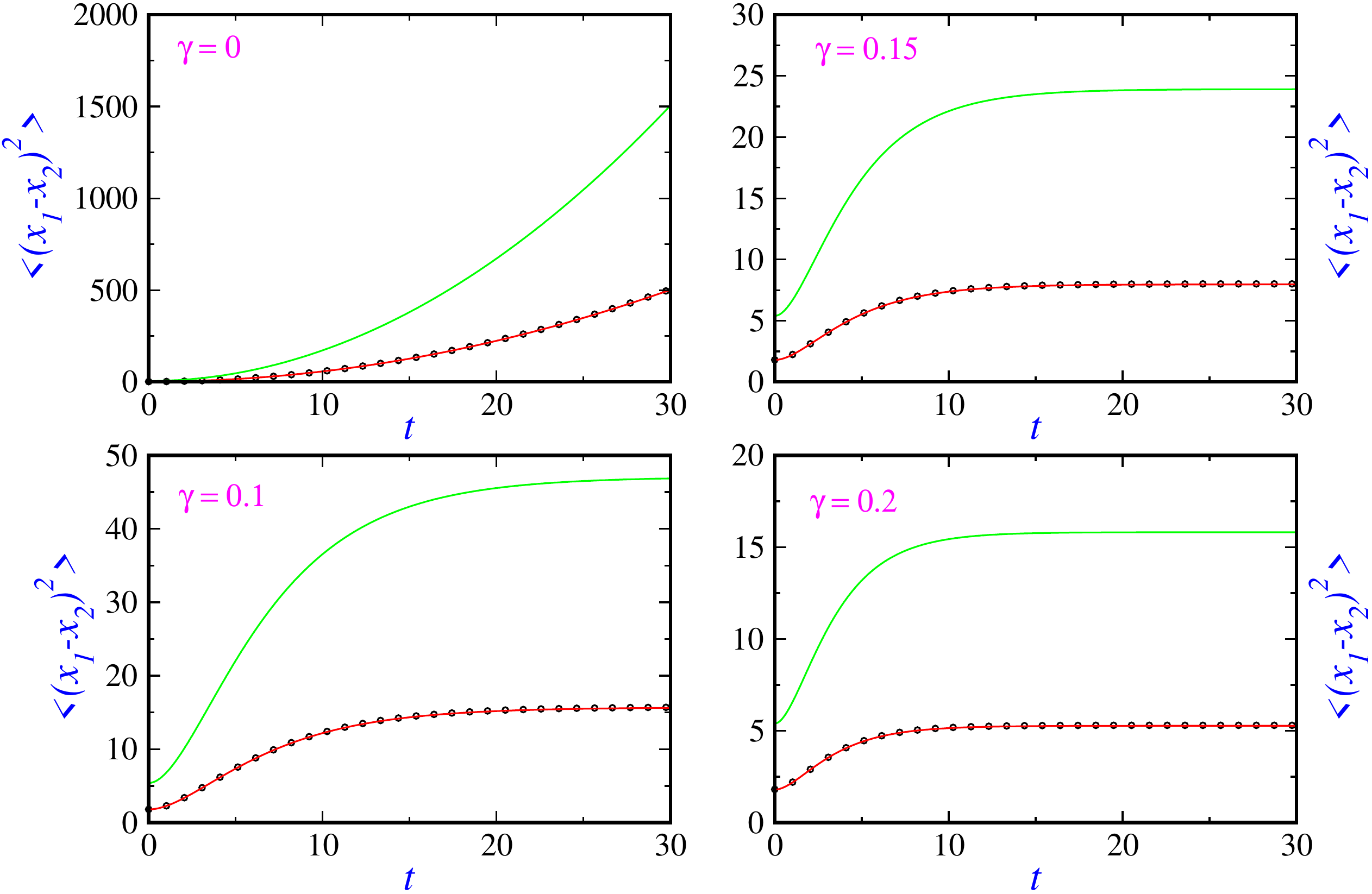}
	\caption{
		Mean square separation (MSS) versus time in the CK framework for the MB (black circle), BE (red curves) and  FD (green curves) statistics 
		and values of friction coefficient in different panels: $ \ga = 0 $ (left top), $ \ga = 0.1 $ (left bottom), $ \ga = 0.15 $ (right top) and $ \ga = 0.2 $ 
		(right bottom). 
		Other parameters of the theory have been fixed as follows: $ \si_0 = 1 $, $ x_0 = 0 $, $ p_0 = 3 $, $ \bar{\si}_0 = 0.9 $ and $\bar{p}_0 = p_0 $.
	}
	\label{fig: MSS_CK} 
\end{figure}
Within the CK framework, in Figure \ref{fig: MSS_CK}  we have plotted MSS versus time for the MB (black circle), BE (red curves) and  
FD (green curves) statistics and values of friction coefficient in different panels: $ \ga = 0 $ (left top), $ \ga = 0.1 $ (left bottom), $ \ga = 0.15 $ 
(right top) and $ \ga = 0.2 $ (right bottom). For the non-dissipative case, the MSS is an increasing function of time. This behavior is more pronounced for
fermions than bosons and distinguishable particles which display the same time behaviour. On the contrary, 
when dissipation is present, a drastic behavior is observed. The MSS is much smaller with friction and an asymptotic value or stationary regime 
seems to be settled. At very long times, $ t \gg \ga^{-1}$, localization effects tend to be important leading to a drastic decreasing of the MSS. 
This quantity becomes ultimately constant due to the fact that the friction force is acting in opposite direction to the motion of particles. At $t=0$, 
the initial MSS is non-zero for the three statistics here considered. The high values reach ed in the friction-free case makes that these initial values seem 
to be zero. Interestingly enough, as far as MSS and single-particle density concern, there is negligible
difference between identical bosons and distinguishable particles. This point has already been reported in the study of two-particle two-slit experiment 
by computing joint detection probability for identical bosons and distinguishable particles in the context of non-dissipative systems \cite{Sancho-EPJD-2014}.

\begin{figure} 
	\centering
	\includegraphics[width=12cm,angle=-0]{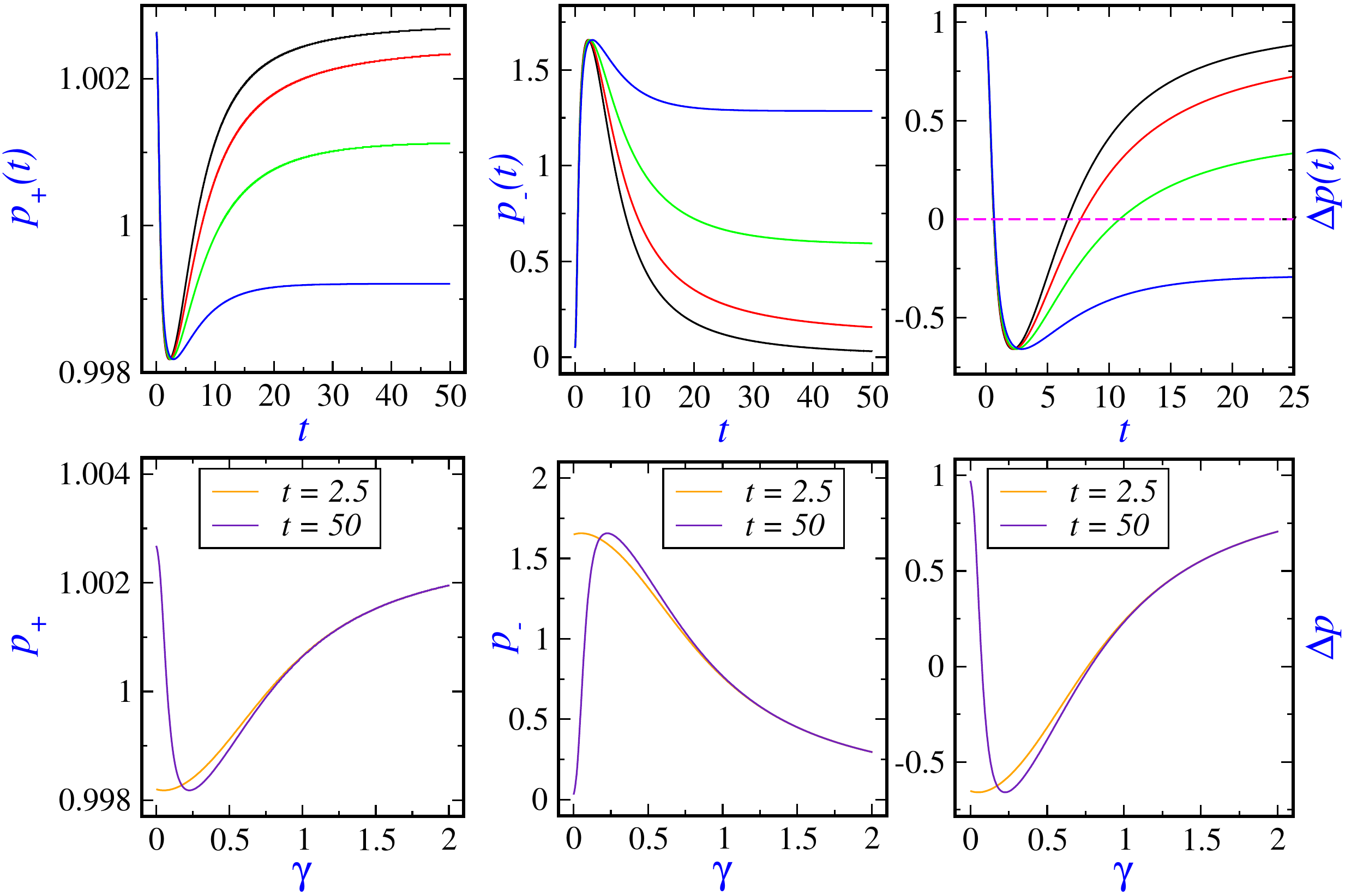}
	\caption{
Relative simultaneous detection probability $ p_+(t) = \frac{ p_{\BE}(t) }{ p_{\MB}(t) } $ (left top plot) for two identical bosons and 
$ p_-(t) = \frac{ p_{\FD}(t) }{ p_{\MB}(t) } $ (middle top plot) for two identical fermions in the CK framework versus time for four different values 
of friction coefficient, $ \ga = 0 $ (black curves), $ \ga = 0.02 $ (red curves), $ \ga = 0.05 $ (green curves) and $ \ga = 0.1 $ (blue curves). 
The difference between the relative joint detection probabilities for bosons and fermions, $ \Delta p(t) = p_+(t) - p_-(t) $, is also plotted in the right 
top panel. In the bottom panels, the same three quantities are plotted versus friction at two fixed times, intermediate time $ t = 2.5 $ (orange curves) 
and stationary time $ t = 50 $ (indigo curves), measured by a detector with a width $2d=2$ located at the origin.
Other parameters have been fixed as follows: $ \si_0 = 1 $, $ x_0 = 0 $, $ p_0 = 3 $,  $ \bar{\si}_0 = 0.9 $ and $\bar{p}_0 = p_0 $. 
	}
	\label{fig: detprob_CK} 
\end{figure}

In Figure \ref{fig: detprob_CK},  the ratio of simultaneous detection probability of indistinguishable particles to the distinguishable ones 
is plotted versus time for bosons $ p_+(t) = \frac{ p_{\BE}(t) }{ p_{\MB}(t) } $ (left top panel) and fermions $ p_-(t) = \frac{ p_{\FD}(t) }
{ p_{\MB}(t) } $ (middle top panel) in the CK framework.
The difference between the relative joint detection probabilities for bosons and fermions, $ \Delta p(t) = p_+(t) - p_-(t) $, is plotted in the right 
top panel.
They are measured by a single extended detector with a width $2d=2$ located at the origin 
for different values of friction coefficient, $ \ga = 0 $ (black curves), $ \ga = 0.02 $ (red curves), $ \ga = 0.05 $ (green curves) and $ \ga = 0.1 $ 
(blue curves). This ratio is around one for bosons meaning that bosons and distinguishable particles have approximately a similar time behaviour.
%
However, for fermions, a quite different behaviour of this ratio is clearly observed. 
For these values of the friction coefficient, the stationary value of the relative detection probability decreases (increases) with friction for bosons (fermions). 
This point is better stressed in the three bottom panels of the same figure where the same three quantities are plotted versus friction 
at two fixed times, intermediate time $ t = 2.5 $ (orange curves) and stationary time $ t = 50 $ (indigo curves), measured by a detector with a 
width $2d=2$ located at the origin.
After our choice of parameters, these plots show that only for $ \ga \leq 0.22 $, the stationary value of detection probability is 
decreasing (increasing) with friction for bosons (fermions). 
The role played by dissipation is to modify the stationary value of the detection probability. The analytical 
form of $p_{\pm}(t)$ is dictated by the (anti)symmetrization of the state, and the intensity by the dissipation.
Furthermore, as the right top panel shows, there is a time interval, increasing with friction, where the detection probability in the detector region is higher for fermions 
than for bosons revealing a sort of fermion-bunching and boson-anti-bunching, just the opposite effect one should expect. 
As the bottom right panel shows, there is a critical value of the friction coefficient $ \ga_c \approx 0.78 $ where for $ \ga > \ga_c $ fermion-bunching is not observed anymore, i.e., $ \Delta p > 0 $.
Moreover, for these values of friction, the detection probability for both times $t=2.5$ and $t=50$ becomes the same revealing that the stationary 
behaviour is seen at times of the order of the relaxation time   $ t \sim 1/ \ga_c $.

%
%
\begin{figure} 
	\centering
	\includegraphics[width=12cm,angle=-0]{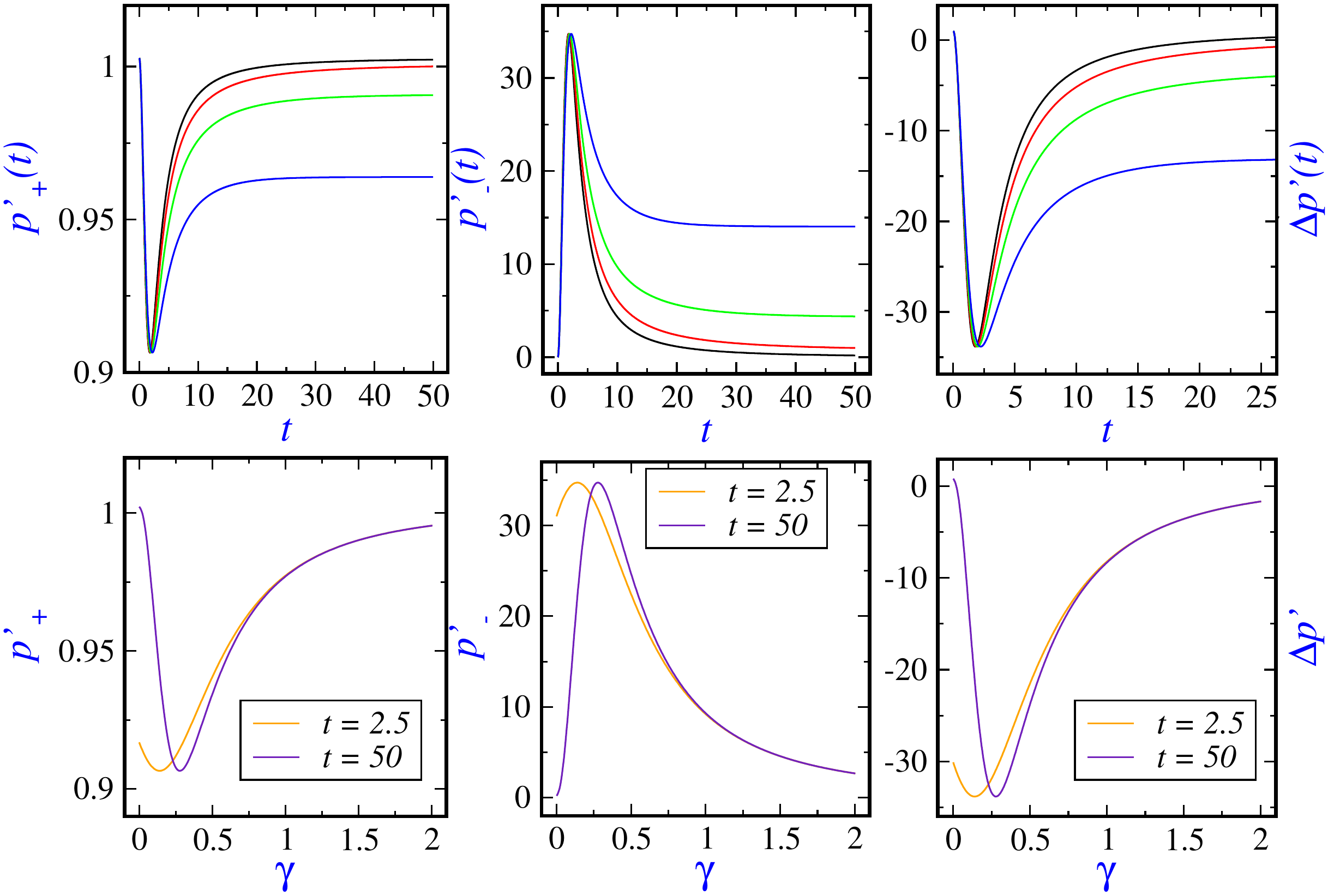}
	\caption{
The same as Figure \ref{fig: detprob_CK}  but for two point detectors. The relative simultaneous detection probability are now given by 
$ p'_+(t) = \frac{ p'_{\BE}(t) }{ p'_{\MB}(t) } $ (two identical bosons) and $ p'_-(t) = \frac{ p'_{\FD}(t) }{ p'_{\MB}(t) } $ (two identical 
fermions) in the CK framework. The detectors are located at $D$ and $-D$ with $D=1$.
	}
	\label{fig: detprob_prime_CK} 
\end{figure}

In Figure \ref{fig: detprob_prime_CK}, the same information is plotted as in Figure \ref{fig: detprob_CK} but for two point detectors located at $D$ and $-D$ with $D=1$.
The same trends are observable here but much more pronounced. The fermion-bunching is clearly enhanced under this new detection scheme.  
Furthermore, in this new scheme, fermion-bunching is seen for all values of friction in the considered interval, i.e., $\Delta p' < 0 $ 
in the whole range of friction for the intermediate time $t=2.5$.

\subsubsection{The Caldeira-Leggett approach}

With the initial Gaussian wave packets given by Eqs. (\ref{eq: psi0}) and (\ref{eq: phi0}), the diagonal elements of one-particle states have 
the form
\begin{numcases}~
\rho_{aa}(x, x, t) = \frac{1}{\sqrt{2\pi} w_t} \exp\left[ - \frac{ (x - x_t)^2}{ 2 w_t^2 }  \right] 
\label{eq: rhoaa}
\\
\rho_{ab}(x, x, t) = \sqrt{ 2 \frac{\si_0 \bar{\si}_0 }{ \si_0^2 + \bar{\si}_0^2 } } 
~ \frac{1}{2 \sqrt{\pi a_2(t)}} \exp \left[ a_0 - \frac{ (x - a_1(t))^2}{ 4 a_2(t) }  \right]
\label{eq: rhoab}
\end{numcases}
under the evolution equation (\ref{eq: CL eq}) with
\begin{numcases}~
w_t = \sqrt{ \si_t^2 + D \frac{ 4 \ga t + 4 e^{-2\ga t} - 3 - e^{-4\ga t} }{ 8 m^2 \ga^3 } } 
\label{eq: wt}
\\
a_0 =  - \frac{\si_0^2 \bar{\si}_0^2 }{ \si_0^2 + \bar{\si}_0^2 }  \frac{(p_0-\bar{p}_0)^2}{\hb^2}
\\
a_1(t) = x_0 + \frac{ p_0 \si_0^2 + \bar{p}_0 \bar{\si}_0^2 }{ m( \si_0^2 + \bar{\si}_0^2 ) } \uptau(t)
+ i \frac{\si_0^2 \bar{\si}_0^2 }{ \si_0^2 + \bar{\si}_0^2 }  \frac{2(p_0-\bar{p}_0)}{\hb}
\\
a_2(t) = \frac{\si_0^2 \bar{\si}_0^2 }{ \si_0^2 + \bar{\si}_0^2 } + \frac{\hb^2 \uptau(t)^2 }{ 4m^2( \si_0^2 + \bar{\si}_0^2 ) } + D \frac{ 4 \ga t + 4 e^{-2\ga t} - 3 - e^{-4\ga t} }{ 16 m^2 \ga^3 } 
- i  \frac{ \hb }{ 2m } \frac{\si_0^2 - \bar{\si}_0^2 }{ \si_0^2 + \bar{\si}_0^2 } \uptau(t)    .
\label{eq: a2}
\end{numcases}
In order to obtain $ \rho_{bb}(x, x, t) $ it suffices to replace $x_t$ by $\bar{x}_t$ in Eq. (\ref{eq: rhoaa}) and $ \si_t $ by $\bar{\si}_t$ in 
Eq. (\ref{eq: wt}).  
In a similar way, $ \rho_{ba}(x, x, t) $ is known from Eq. (\ref{eq: rhoab}) by interchanging $ \si_0 \leftrightarrow \bar{\si}_0 $. 
The temperature appears
in the expressions for the widths through the diffusion constant, $D$.


From the above relations and their equivalent one in Eq. (\ref{eq: mss_CL}), one has that
\begin{numcases}~
\int_{-\infty}^{\infty} dx ~ \rho_{ab}(x, x, t) = \sqrt{ \frac{2 \si_0 \bar{\si}_0 }{ \si_0^2 + \bar{\si}_0^2 } } ~ e^{a_0} 
\\
\la x \ra_{ab} =  \sqrt{ \frac{2 \si_0 \bar{\si}_0 }{ \si_0^2 + \bar{\si}_0^2 } } ~ a_1(t) ~ e^{a_0} 
\\
\la x^2 \ra_{ab} =  \sqrt{ \frac{2 \si_0 \bar{\si}_0 }{ \si_0^2 + \bar{\si}_0^2 } } ~ [ a_1(t)^2 + 2 a_2(t) ] ~ e^{a_0} 
\end{numcases}
leading to the MSS in the CL framework.


Analogously, by using the following expressions
\begin{numcases}~
\int_{-d}^{d} dx ~ \rho_{aa}(x, x, t) =  
\frac{1}{2} \left\{ \erf\left( \frac{x_t + d }{ \sqrt{2} w_t }  \right) - 
\erf\left( \frac{x_t - d }{ \sqrt{2} w_t }  \right) \right\}
\\
\int_{-d}^{d} dx ~ \rho_{ab}(x, x, t) = \sqrt{ \frac{2 \si_0 \bar{\si}_0 }{ \si_0^2 + \bar{\si}_0^2 } } ~ e^{a_0} 
~ \frac{1}{2} \left\{ \erf\left( \frac{a_1(t) + d }{ 2 \sqrt{a_2(t)} }  \right) - 
\erf\left( \frac{a_1(t) - d }{ 2 \sqrt{a_2(t)} }  \right) \right\}
\end{numcases}
and similar ones for the integration of $ \rho_{bb}(x, x, t) $ and $ \rho_{ba}(x, x, t) $ in Eq. (\ref{eq: detprob_CL}), one can calculate the 
joint detection probability of both identical particles by an extended detector with width $2d$ in the CL approach. 

 \begin{figure} 
 	\centering
 	\includegraphics[width=12cm,angle=-0]{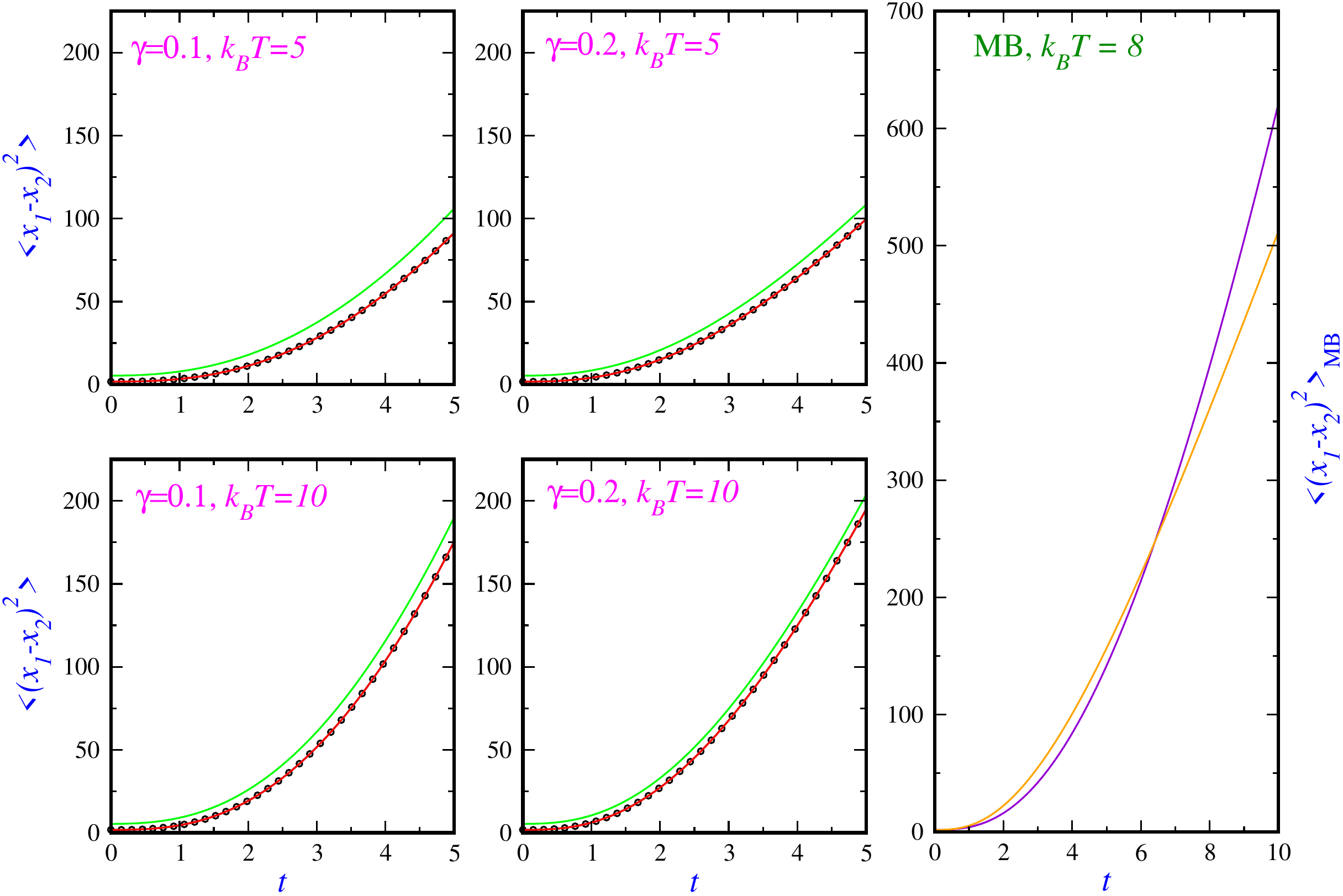}
 	\caption{
 		Mean square separation versus time in the CL framework for different statistics, MB (black circles), BE (red curves), FD (green curves) 
 		for different values of friction, $\ga = 0.1$ (left panels) and $\ga = 0.2$ (middle panels) for
 		$ k_B T = 5 $ (top panels) and $ k_B T = 10 $ (bottom panels). 
 		Right panel depicts MSS for the MB statistics for $ k_B T = 8 $ and two friction values, $\ga = 0.1$ (violet) and $\ga = 0.2$  (orange).
 		Other parameters have been fixed as follows: $ \si_0 = 1 $, $ x_0 = 0 $, $ p_0 = 3 $, $ \bar{\si}_0 = 0.9 $ and $\bar{p}_0 = p_0 $.
 	}
 	\label{fig: MSS_CL} 
 \end{figure}

In this approach, the is displayed versus time in Figure \ref{fig: MSS_CL} for different statistics: MB (black circles), BE (red curves), 
FD (green curves) for different values of friction, $\ga = 0.1$ (left panels) and $\ga = 0.2$ (middle panels) with
$ k_B T = 5 $ (top panels) and $ k_B T = 10 $ (bottom panels). Right panel depicts the MSS for MB statistics at $ k_B T = 8 $ for $\ga = 0.1$ 
(violet) and $\ga = 0.2$  (orange). 
It is clear that the MSS is higher for identical fermions than identical bosons. This separation for bosons is slightly lower than for distinguishable 
particles although this  is not clearly seen due to the scale of the plots.
This quantity also increases with time and temperature for a given friction with no asymptotic value. 
After the right panel, the MSS for the MB statistics displays a nearly linear behavior with time where the slope depends on the temperature and friction.  
Although all terms contributing to the MSS in Eq. (\ref{eq: mss_CL_MB}) reduce with friction for a given time, the rate of such a reduction is different for each term.
This behavior is also seen for different statistics.
%
The time dependent variation is quite different to the CK approach. This is due to the presence of thermal effects where the 
corresponding temperature makes the widths of the wave packets increase. In a certain sense, temperature and friction play an opposite 
effect on the width of the wave packet. 


%
 \begin{figure} 
 	\centering
 	\includegraphics[width=12cm,angle=-0]{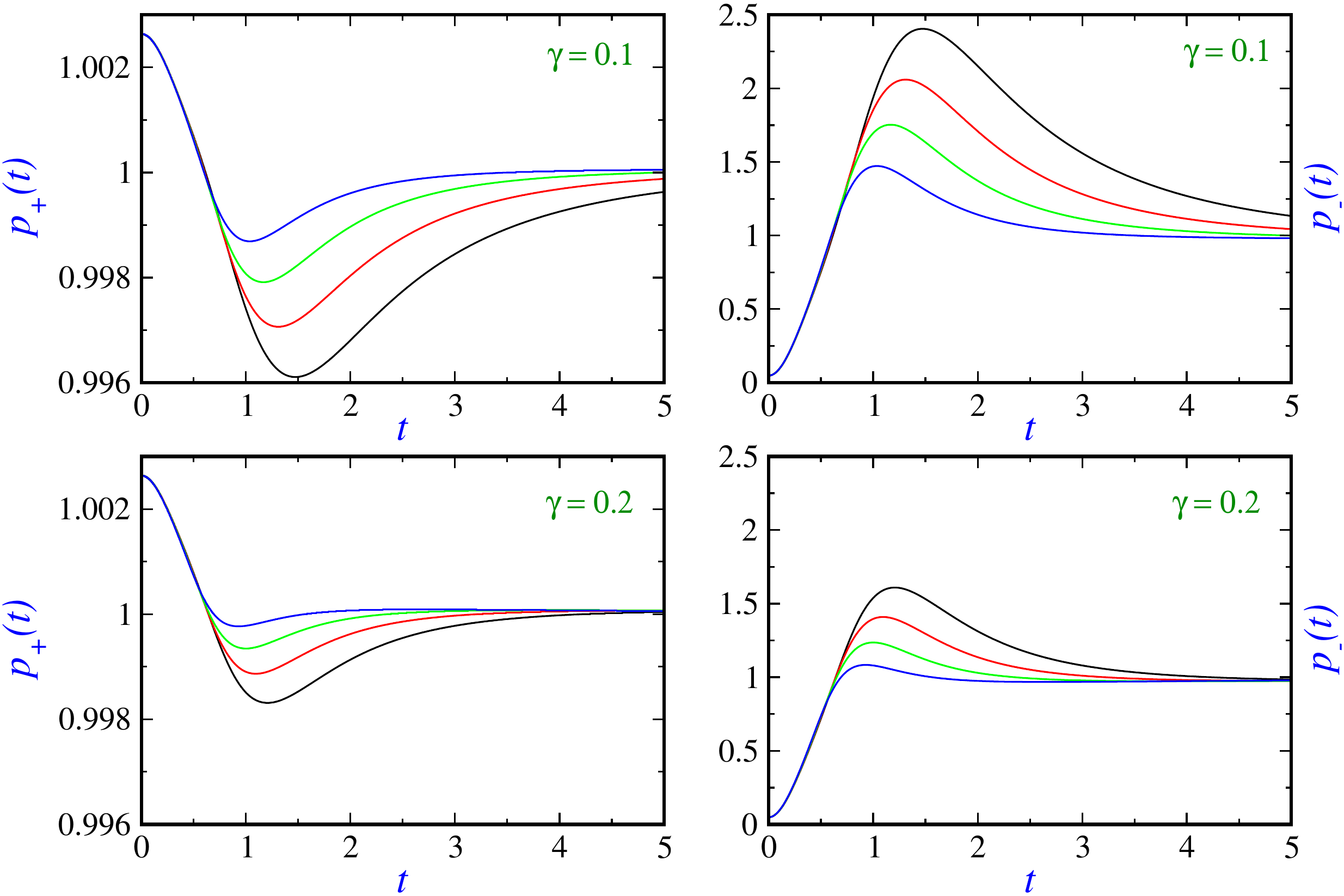}
 	\caption{
 		Relative simultaneous detection probability $ p_+(t) = \frac{ p_{\BE}(t) }{ p_{\MB}(t) } $ (left plots) for two identical bosons and 
 		$ p_-(t) = \frac{ p_{\FD}(t) }{ p_{\MB}(t) } $ (right plots) for two identical fermions in the CL framework, measured by a detector 
 		with a width $2d=2$ located at the origin, for $\ga = 0.1$ (top plots) and $\ga = 0.2$ (bottom plots) for different values of temperature: 
 		$ k_B T = 5 $ (black curves), $ k_B T = 7 $ (red curves), $ k_B T = 10 $ (green curves) and $ k_B T = 15 $ (blue curves). Other parameters have been fixed as follows: $ \si_0 = 1 $, $ x_0 = 0 $, $ p_0 = 3 $, $ \bar{\si}_0 = 0.9 $ and $\bar{p}_0 = p_0 $.
 	}
 	\label{fig: detprob_CL} 
 \end{figure}

In Figure  \ref{fig: detprob_CL} ,the ratio of simultaneous detection probability of indistinguishable particles to the distinguishable ones 
is plotted versus time for bosons $ p_+(t) = \frac{ p_{\BE}(t) }{ p_{\MB}(t) } $ (left panels) and $ p_-(t) = \frac{ p_{\FD}(t) }{ p_{\MB}(t) } $
(right panels) for fermions, measured by an extended detector with a width $2d=2$ located at the origin, for $\ga = 0.1$ (top plots) and $\ga = 0.2$ 
(bottom plots) for different values of temperature: $ k_B T = 5 $ (black curves), $ k_B T = 7 $ (red curves), $ k_B T = 10 $ (green curves) and $ k_B T = 15 $ 
(blue curves). The general behavior for both cases is similar to that  found within the CK approach. For bosons, the temperature plays a major role 
than friction but the ratios are always around one. In particular, by increasing the temperature, the ratio is approaching to one. For fermions,
the ratios become greater than one, after a while, but decrease with temperature. Again, the bunching and anti-bunching behavior is observed for fermions and 
bosons, respectively. In any case, interestingly enough, the two kind of ratios ultimately reach one for all temperatures and frictions 
analyzed. Thus, decoherence process, loss of being indistinguishable, is settled gradually by increasing friction and temperature. 
The symmetry of the total wave function is not so important under these conditions. The property of being distinguishable is emerging gradually.

\begin{figure} 
\centering
\includegraphics[width=12cm,angle=-0]{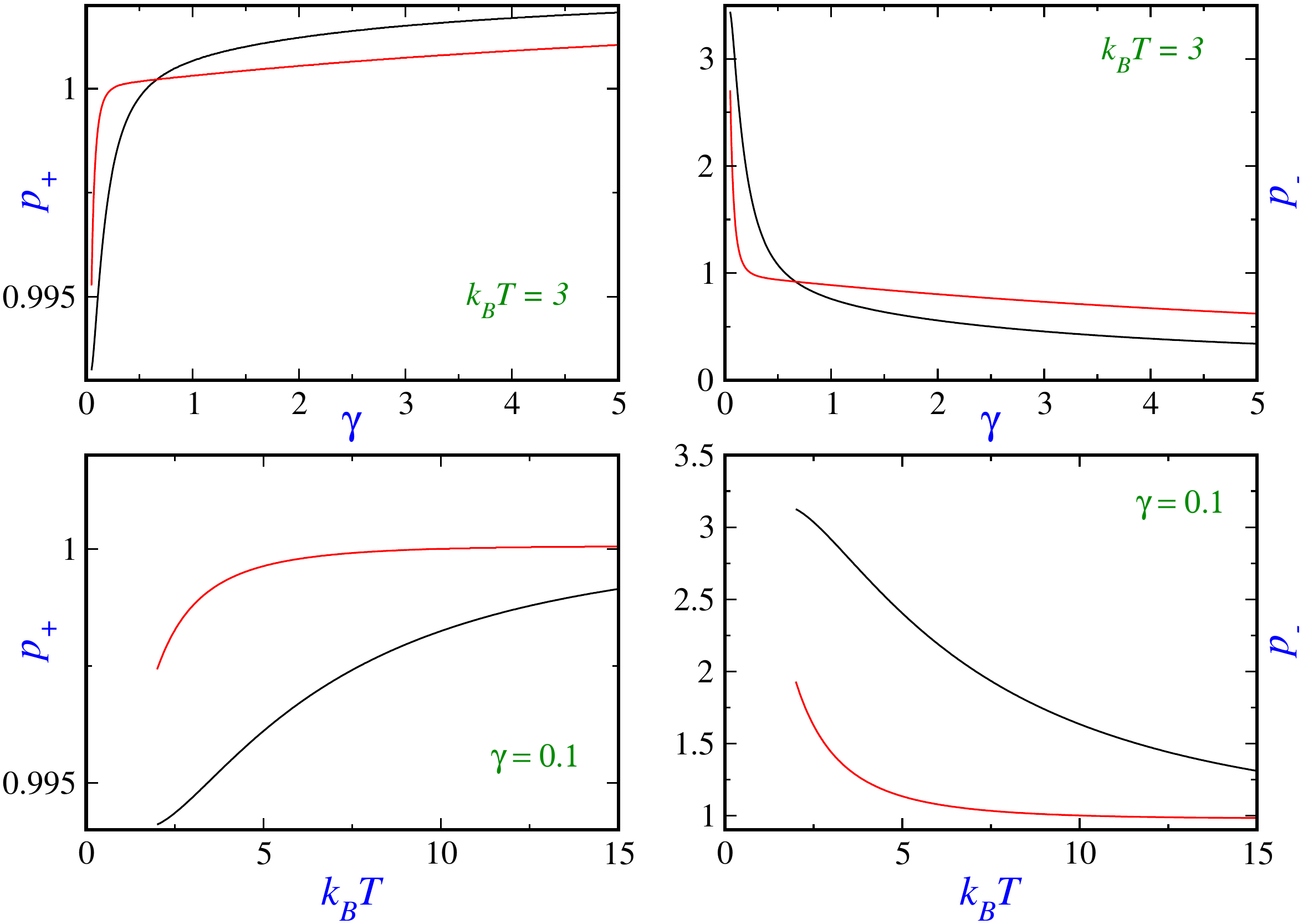}
\caption{
Relative simultaneous detection probability in the CL framework at two fixed times $ t = 1.5 $ (black curves) and $ t = 5 $ (red curves) versus 
friction coefficient for $ k_B T = 3 $ (top plots) and versus temperature for $\ga = 0.1$ (bottom plots), measured by a detector with a width 
$2d=2$ located at the origin. Other parameters have been fixed as follows: $ \si_0 = 1 $, $ x_0 = 0 $, $ p_0 = 3 $, $ \bar{\si}_0 = 0.9 $ and $\bar{p}_0 = p_0 $.
}
\label{fig: detprob_CL2} 
 \end{figure}

It is then clear that similar behaviors are observed in both frameworks concerning detection probabilities. The extra difference is that when 
adding temperature, the exchange effects become less important at high temperatures leading to the same behaviour for bosons and 
fermions in this regime. This is better observed in Figure \ref{fig: detprob_CL2}, where the relative
simultaneous detection probability  in the CL framework at two fixed times $ t = 1.5 $ (black curves) and $ t = 5 $ (red curves) versus friction 
coefficient for $ k_B T = 3 $ (top panels) and versus temperature for $\ga = 0.1$ (bottom panels), measured by a detector with a width $2d=2$ 
located at the origin, are displayed.   

\subsection{The two-particle two-slit experiment: the CK approach} \label{sec: tp-tl}

\begin{figure} 
\centering
\includegraphics[width=8cm,angle=-0]{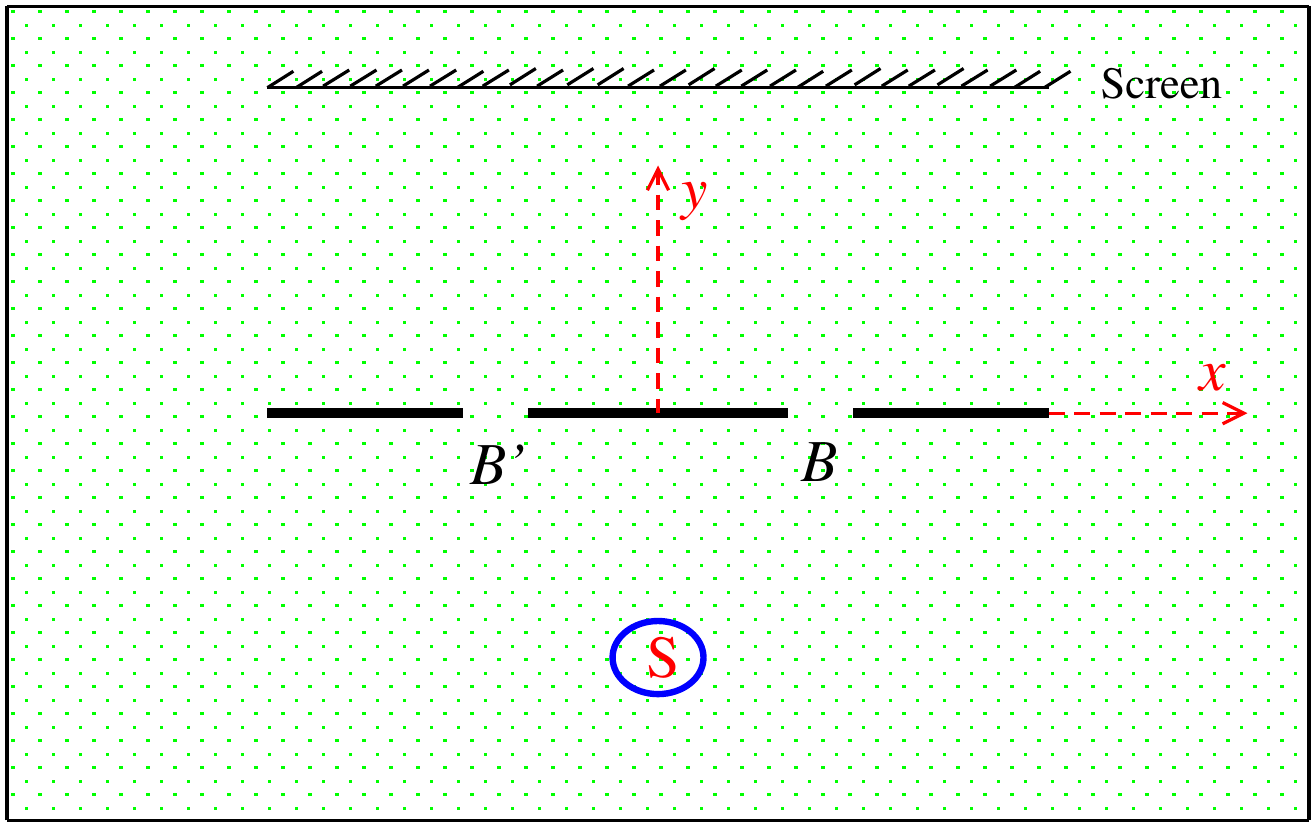}
\caption{
Particles are emitted by pairs from the source S, pass through the two slits $B$ and $B'$ and arrive at the screen.
}
\label{fig: setup} 
\end{figure}

The problem of the two-particle two-slit experiment has been recently studied by Sancho \cite{Sancho-EPJD-2014} for conservative systems. 
Here, we extend this study to dissipative dynamics in the CK approach.
We consider a two-slit interference experiment when the source emits particles by pairs. As is shown in Figure \ref{fig: setup}, 
the two slits are denoted by $B$ and $B'$ located symmetrically at the points $(\pm X, 0)$ and have the same width $w$. Gaussian slits 
are again assumed. Detectors measure the joint patterns by counting simultaneous arrivals. One-particle states are given by  the wave functions $\psi$ and $\phi$. 

Particles are produced in a source located on the negative $y-$axis in a product state which has a Gaussian shape with zero kick momentum in the $x-$direction but plane wave in $y-$direction,
\begin{eqnarray} \label{eq: in_wf}
\psi_0(x, y) &=& A \frac{1}{(2 \pi \si_0^2)^{1/4}} \exp \left[-\frac{x^2}{4\si_0^2} + i k y \right]
\end{eqnarray}
with $\si_0 = \hb / 2\si_p$, $\si_p$ being the momentum width, along $x-$axis, of the wave function and constant $A$.
Then, this wave function propagating freely arrives at time $ t_0 = \hb k / m $ to both slits. Thus, for $ t< t_0 $, the wave function is written as
\begin{eqnarray} \label{eq: wf before t_0}
\psi(x, y, t) &=&
A \frac{1}{(2\pi s_t^2)^{1/4}} \exp\left[- \frac{x^2}{ 4 \si_0 s_t } + i k y - i \frac{E t}{\hb}  \right] , \qquad t<t_0 
\end{eqnarray}
where $ E = \hb^2 k^2 / 2 m $ and $s_t$ is the complex width of the wave function in the $x-$direction given by Eq. (\ref{eq: st}). 
Here, we have assumed that the friction force acts along the $x-$axis only.
Since the motion in the $y-$direction is described by plane waves, in the following we ignore the motion in this direction and consider only 
the dynamics along the plane of slits, i.e, the $x-$direction. 

By using the Gaussian slit approximation, the single particle wave function corresponding to the right slit is given by
\begin{eqnarray} \label{eq: wf after t_0}
\psi_B(x, t) &=&
N \int_{-\infty}^{\infty} dx'~e^{-(x'-X)^2/2w^2} G(x, t; x', t_0) \psi(x', t_0)  , \qquad t>t_0
\end{eqnarray}
where $N$ is the normalization constant taken to be real, the first factor in the integrand is the weight function corresponding to the Gaussian 
slit approximation with $w$ being the width of the slit, and the second factor is the free particle propagator in the CK approach given 
by \cite{MoMi-JPC-2018, Mo-LNC-1978}
\begin{eqnarray} \label{eq: propagator}
G(x, t; x', t_0) &=& \sqrt{ \frac{m}{2\pi i \hb ~ (\uptau(t)-\uptau(t_0))} } \exp \left[ \frac{im (x-x')^2}{2\hb ~ (\uptau(t)-\uptau(t_0))}  \right] .
\end{eqnarray}
By carrying out the corresponding integral, we have that
\begin{eqnarray} \label{eq: wft>t0}
\psi_B(x, t) &=& N \frac{1}{(2\pi s_{t_0}^2)^{1/4}} 
\left\{ 1 + \frac{i\hb}{m} \left( \frac{1}{w^2} + \frac{1}{2 \si_0  s_{t_0}}  \right) (\uptau(t)-\uptau(t_0)) \right\}^{-1/2}
~ e^{- c_2(t) x^2 + c_1(t) x + c_0(t)}
\end{eqnarray}
where the following abbreviations are used
\begin{numcases}~
s_{t_0} = s(t_0) \\
\si_{t_0} = \si(t_0)  
\end{numcases}
with $\si_t$ given by Eq. (\ref{eq: sigmat}) and
\begin{eqnarray} \label{eq: normalization}
N &=& \left( \frac{ w^2 + 2 \si_{t_0}^2 }{ w^2 } \right)^{1/4} \exp\left[ \frac{ X^2 }{ 2 w^2 + 4 \si_{t_0}^2 } \right] 
\end{eqnarray}
where
\begin{numcases}~
c_0(t) = - \frac{ 2m\si_0 s_{t_0} + i \hb (\uptau(t)-\uptau(t_0)) }{ 4 m w^2 \si_0 s_{t_0}  + 2 i \hb ( w^2 +2 \si_0 s_{t_0} ) (\uptau(t)-\uptau(t_0)) } X^2 \\
c_1(t) = \frac{ 4 m\si_0 s_{t_0} }{ 4 m w^2 \si_0 s_{t_0} + 2 i \hb ( w^2 +2 \si_0 s_{t_0} ) (\uptau(t)-\uptau(t_0)) } X  \\
c_2(t) = \frac{ m( w^2 + 2\si_0 s_{t_0}) }{ 4 m w^2 \si_0 s_{t_0} + 2 i \hb ( w^2 +2 \si_0 s_{t_0} ) (\uptau(t)-\uptau(t_0)) }  .
\end{numcases}

The wave function $\psi_{B'}$ is given by Eq. (\ref{eq: wft>t0}) by replacing $X$ by $-X$. The wave functions $\phi_B$ and $\phi_{B'}$ 
are obtained from $\psi_B$ and $\psi_{B'}$ by replacing $ \si_0 \rightarrow \bar{\si}_0 $, $ s_t \rightarrow \bar{s}_t$ ($\bar{s}_t$ being the complex with of 
$\phi$) and  $ \si_t \rightarrow \bar{\si}_t $ ($\bar{\si}_t$ being the with of $\phi$).
\begin{figure} 
	\centering
	\includegraphics[width=12cm,angle=-0]{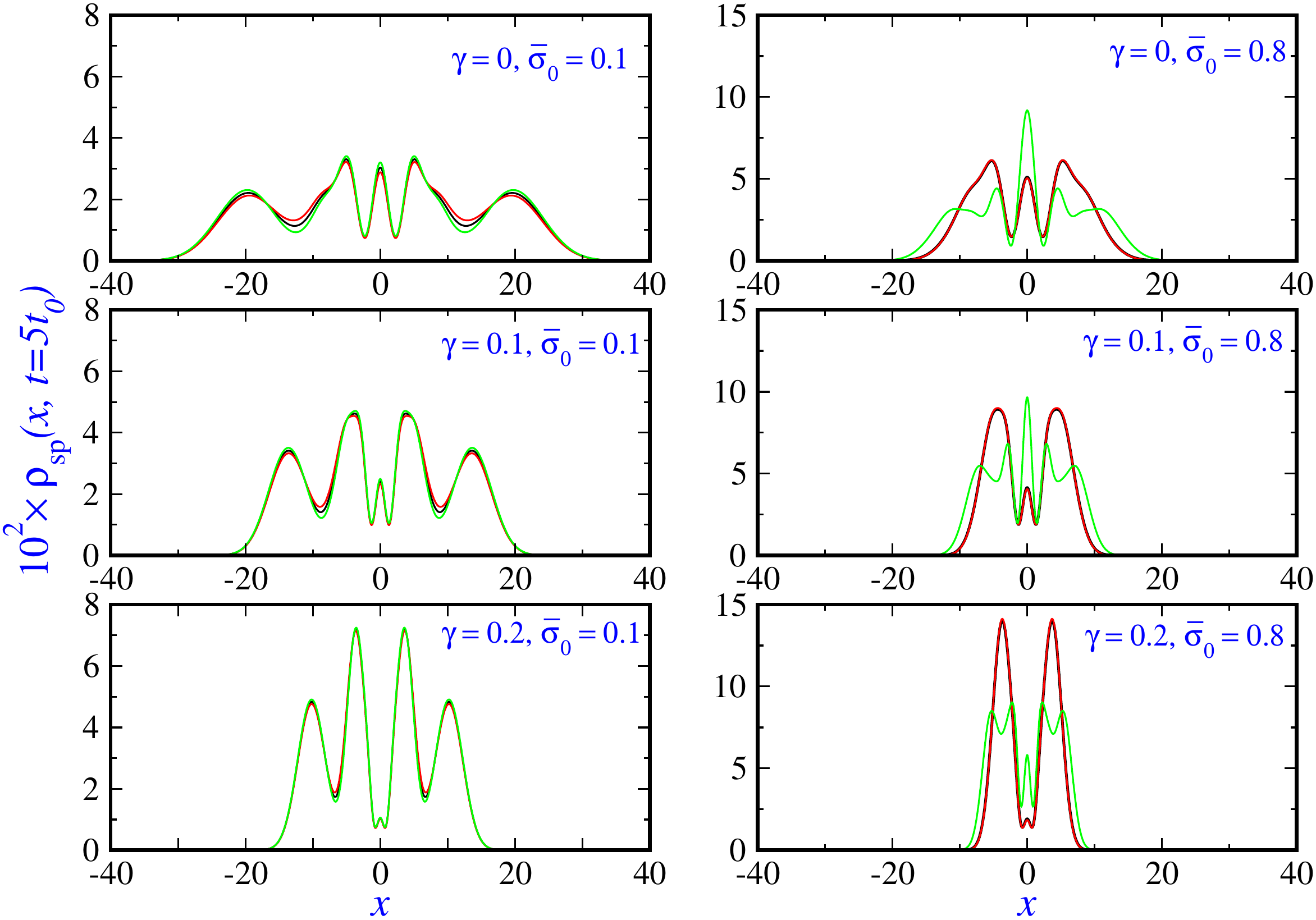}
	\caption{
Scaled single-particle probability density $ \rho_{\text{sp}}(x, t) $ at time $ t = 5 t_0 $ versus space coordinate $x$ for different widths of the one-particle state $\phi(x, t)$, $ \bar{\si}_0 = 0.1 $(left column) and $ \bar{\si}_0 = 0.8 $ (right column) for MB statistics (black curves), BE statistics (red curves) and FD statistics (blue curves) and for friction values $ \ga = 0 $ (top panels), 
$ \ga = 0.1 $ (middle panels) and $ \ga = 0.2 $ (bottom panels).
For numerical calculations we have used $ w = 1 $, $ \si_0 = 0.9 $ and $ t_0 = 1 $.
	}
	\label{fig: rhosp} 
\end{figure}
\begin{figure} 
	\centering
	\includegraphics[width=12cm,angle=-0]{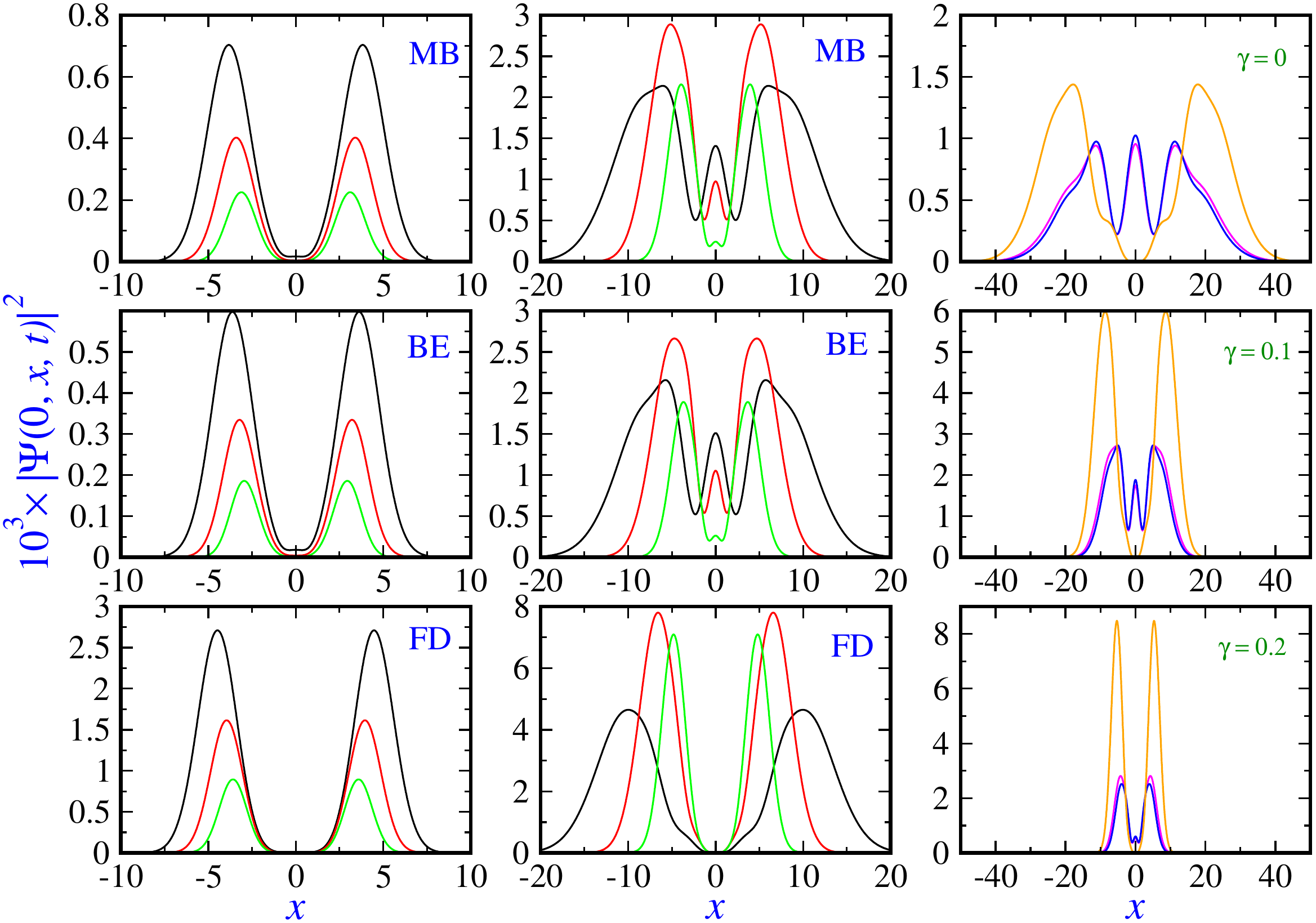}
	\caption{
Scaled joint detection probability at different times $ t = 2 t_0 $ (left column), $ t = 5 t_0 $ (middle column)   
versus the position of moving detector for different statistics and different values of friction coefficient. In the left and middle columns, black 
curves correspond to $ \ga = 0 $, red curves correspond to $ \ga = 0.1 $ while green ones correspond to $ \ga = 0.2 $ for MB statistics (top panels), 
BE statistics (middle panels) and FD statistics (bottom panels). In the right column, the same quantity is plotted but at time $t = 10 t_0$ for the 
MB statistics (magenta curves), BE statistics (blue curves) and FD statistics (orange curves) and friction values $ \ga = 0 $ (top panel), 
$ \ga = 0.1 $ (middle panel) and $ \ga = 0.2 $ (bottom panel). 
For numerical calculations we have used $ w = 1 $, $ \si_0 = 0.9 $, $ \bar{\si}_0 = 0.7 $, $ X = 4 $ and $ t_0 = 1 $.
	}
	\label{fig: rho2p} 
\end{figure}
In our two-particle double-slit experiment, the total wave functions for identical particles are given by Eq. (\ref{eq: 2p-psit}) where
\begin{numcases}~
\psi(x, t) = N_{\psi} (\psi_{B}(x, t) + \psi_{B'}(x, t)) \\
\phi(x, t) = N_{\phi} (\phi_{B}(x, t) + \phi_{B'}(x, t))   .
\end{numcases}
%
Apart from a phase factor, the normalization constants are expressed as
\begin{numcases}~
N_{\psi} = [ 2 ( 1 + \re\{\la \psi_{B} | \psi_{B'} \ra\} )  ]^{-1/2} \\
N_{\phi} = [ 2 ( 1 + \re\{\la \phi_{B} | \phi_{B'} \ra\} )  ]^{-1/2}   .
\end{numcases}
Note that according to Eq. (\ref{eq: interference}), interference terms, inner products of one-particle states, and thus normalization constants 
are independent on time.

Now, the single-particle density $\rho_{\text{sp}}(x, t)$, that is, the probability density for finding a particle at time $t$ at $x$, irrespective 
of the position of the other particle of the pair, evaluated from Eq. (\ref{eq: rhosp}), and 
the joint detection probability with two detectors, one fixed in the origin and the other moving along the slits' plane can be computed, that is, 
$ | \Psi(x_1=0, x_2 = x, t) |^2 $ for different statistics. For numerical calculations, the following parameters $ X = 4 $, $ w = 1 $, $ \si_0 = 0.9 $, 
$ \bar{\si}_0 = 0.7 $, otherwise stated, and $ t_0 = 1 $ are used.
%
%
In Figure \ref{fig: rhosp}, we have plotted the single particle probability density at time $t=5t_0$ versus space coordinate $x$ for different
statistics and friction values. Two different values of widths of the one-particle state $\phi(x, t)$ are used, fixing the width of the other one-particle state, 
that is, $\psi(x, t)$.
As this figure shows, when there is little overlap between one-particle states, $\si_0 = 0.9 $ and $\bar{\si}_0 = 0.1 $ all kinds of particles behave similarly because  the exchange effects are not so important and only introduce small differences among the three types of
particles. The differences among the three curves decrease when dissipation increases. However, when considerable overlapping is present, 
$\si_0 = 0.9 $ and $\bar{\si}_0 = 0.8 $, fermions behave completely different from bosons which themselves behave quite similar to 
distinguishable particles, a result which has already been noticed in the context of non-dissipative systems \cite{Sancho-EPJD-2014}. 

In Figure \ref{fig: rho2p} the joint detection probability versus the position of the moving detector is plotted at two times, $ t = 2 t_0 $ 
(left columns) and $ t = 5 t_0 $ (middle columns), for the MB (top panels), BE (middle panels) and FD statistics (bottom panels)
and three values of the friction coefficient $ \ga = 0 $ (black curves), $ \ga = 0.1 $ (red curves) and $ \ga = 0.2 $ (green curves). 
In the right column, the same quantity is plotted but at time $t = 10 t_0$ for the MB statistics (magenta curves), BE statistics (blue curves) 
and FD statistics (orange curves) and friction values $ \ga = 0 $ (top panel), $ \ga = 0.1 $ (middle panel) and $ \ga = 0.2 $ (bottom panel). 
At short times (left panels), the joint detection probability starts spreading, the interference being not so important yet. With friction, the intensities are 
decreasing and, for fermions, the highest intensities are reached. At intermediate times (middle panels) $t=5t_0$, the interference process is already 
important for the MB and BE statistics whereas, for the FD one, a higher spreading is clearly observed. 
Finally, at $ t = 10 t_0 $ (right panels), a similar behavior is observed. The different patterns for fermions are still formed by two lobes splitting 
apart each other, emphasizing the anti-bunching property of these particles even with friction. Furthermore, the lobes are sharper with increasing friction.
Thus, for the set of the parameters chosen, bosons behave like distinguishable particles while fermions have a complete different behavior, 
reflecting in a certain sense the bunching and anti-bunching properties of bosons and fermions, respectively. As expected, the interference is decreasing with 
friction for bosons and distinguishable particles.
In the three panels of the right column, the spreading is drastically reduced by friction but, on the contrary, the intensity peaks are higher with friction.  We attribute these behaviours to a manifestation of  localization effects due to dissipation, as has been already mentioned above.
%

\section{Concluding remarks} 

The importance of friction and temperature leading to the decoherence process in open quantum systems is very well known. In this work, we have 
analyzed their mutual influence for non-interacting, distinguishable and indistinguishable particles (bosons and fermions) by considering the 
interference and diffraction patterns by one slit within the CK and CL approaches and two Gaussian slits only in the CK one  taking the one-particle states with considerable overlap. 
The mean square separation, computed only for the one Gaussian slit problem in both CK and CL frameworks, is always greater for fermions 
than for bosons. The counter intuitive bunching and anti-bunching effects described in \cite{MaGr-EPJD-2014} for fermions and bosons, respectively,  are observed through the 
detection probability. 
Our work notoriously extends the scope of this unusual behavior by showing its presence (i) in scenarios where there are not zeros in the single-particle wavefunctions, and (ii) in one- and two-detector schemes (in \cite{MaGr-EPJD-2014} is only for the second type).
The time dependent probability tends to be the same for bosons and distinguishable particles but quite different for fermions. 
Decoherence process, loss of being indistinguishable, is settled gradually with time by increasing friction and temperature.
%
%
In the two slits case, by computing single-particle detection probability for different particles in the context of the CK model, we 
have observed that all kinds of particles behave almost similarly for low-overlapping one-particle states. This is due to the fact that for low values 
of overlapping, the exchange effects are not so important and only introduce small differences among the three types of
particles. These differences decrease as friction increases. On the contrary, for considerable overlapping where exchange effects are important, 
fermions behave completely different from bosons which themselves behave like distinguishable particles. These findings (i) show that in the 
regime considered the overlapping degree (and not the relaxation constant) is the fundamental parameter in the problem, and (ii) provide a 
confirmation of the results in \cite{Sancho-EPJD-2014} but for open systems.
This work should be seen as a good starting point to study optical properties of matter waves under the presence of friction and 
temperature when considering non-interacting identical particles governed by the two quantum statistics. In particular, the  Talbot effect 
leading to  quantum carpets could be a good candidate. Furthermore, how the quantum statistics can influence 
the dissipative quantum backflow is another aspect to take into account to see the bunching and anti-bunching effects here described. Obviously, 
the list of interesting topics to be analyzed  and discussed in a future work within this context  is enormous. 

\vspace{1cm}
\noindent
{\bf Acknowledgement}
\vspace{1cm}

SVM acknowledges support from the University of Qom and SMA support from 
the Ministerio de Ciencia, Innovaci\'on y Universidades (Spain) under the
Project FIS2017-83473-C2-1-P.

\newpage


\appendix

\section{Two-particle Schr\"{o}dinger-Langevin equation}

In this appendix we construct the two-particle Schr\"{o}dinger-Langevin equation and show that the symmetrization procedure does  not work.

Suppose $\psi(x, t)$ and $\phi(x, t)$ are two single-particle wave functions satisfying the single-particle Schr\"{o}dinger-Langevin equations,
\begin{numcases}~
i \hb \frac{\pa }{\pa t} \psi(x_1, t) = \bigg[ - \frac{\hb^2}{2m} 
\frac{\pa^2}{\pa x_1^2} + V(x_1) 
+ \frac{\hb \ga}{2i} \left( \ln \frac{\psi(x_1, t) }{\psi^*(x_1, t) } - 
\left\langle \psi \bigg|  \ln \frac{\psi }{\psi^* } \bigg| \psi \right\rangle  \right)
\bigg] \psi(x_1, t) \label{eq: sch_a1}
\\
i \hb \frac{\pa }{\pa t} \phi(x_2, t) = \bigg[ - \frac{\hb^2}{2m} 
\frac{\pa^2}{\pa x_2^2} + V(x_2) 
+ \frac{\hb \ga}{2i} \left( \ln \frac{\phi(x_2, t) }{\phi^*(x_2, t) } -
\left\langle \phi \bigg|  \ln \frac{\phi }{\phi^* } \bigg| \phi \right\rangle  \right)
  \bigg] \phi(x_2, t) \label{eq: sch_b2}
\end{numcases}
Let us now consider a system composed of two identical particles. In order to derive a wave equation for such a system,
we multiply (\ref{eq: sch_a1}) by $\phi(x_2, t)$ and (\ref{eq: sch_b2}) by $\psi(x_1, t)$ and then add the resulting equations. 
In this way, we obtain 
\begin{eqnarray} \label{eq: sch_1}
i \hb \frac{\pa }{\pa t} [ \psi(x_1, t) \phi(x_2, t) ] &=& H(x_1, x_2; t) \psi(x_1, t) \phi(x_2, t)
\end{eqnarray}
where,
\begin{eqnarray} \label{eq: 2parham}
H(\psi(x_1),\phi(x_2);t) &=& \bigg[ - \frac{\hb^2}{2m} 
\left( \frac{\pa^2}{\pa x_1^2} + \frac{\pa^2}{\pa x_2^2}\right) + V(x_1) + V(x_2)
\nonumber \\
&+& \frac{\hb \ga}{2i} \left( \ln \frac{\psi(x_1, t) \phi(x_2, t)}{\psi^*(x_1, t) \phi^*(x_2, t)} 
- \left\langle \psi \bigg|  \ln \frac{\psi }{\psi^* } \bigg| \psi \right\rangle
- \left\langle \phi \bigg|  \ln \frac{\phi }{\phi^* } \bigg| \phi \right\rangle
  \right)
  \bigg] \psi(x_1, t) \phi(x_2, t)
\nonumber \\
\end{eqnarray}
is equivalent to a Hamiltonian for the two-particle system which clearly shows that is not symmetric under the exchange of particles.
One can try it by hand and write,
\begin{eqnarray} \label{eq: sch_1*}
i \hb \frac{\pa }{\pa t} \psi(x_1, t) \phi(x_2, t) &=& \bigg[ - \frac{\hb^2}{2m} 
\left( \frac{\pa^2}{\pa x_1^2} + \frac{\pa^2}{\pa x_2^2}\right) + V(x_1) + V(x_2)
\nonumber \\
&+& \frac{\hb \ga}{2i} \left( 
\frac{1}{2}\ln \frac{\psi(x_1, t) \phi(x_2, t)}{\psi^*(x_1, t) \phi^*(x_2, t)}
+ \frac{1}{2} \ln \frac{\psi(x_2, t) \phi(x_1, t)}{\psi^*(x_2, t) \phi^*(x_1, t)}
- \left\langle \psi \bigg|  \ln \frac{\psi }{\psi^* } \bigg| \psi \right\rangle
- \left\langle \phi \bigg|  \ln \frac{\phi }{\phi^* } \bigg| \phi \right\rangle
  \right)
  \bigg] 
\nonumber \\  
&\times&  \psi(x_1, t) \phi(x_2, t)
\nonumber \\
\end{eqnarray}
This equation can be re-written as
\begin{eqnarray} \label{eq: sch_13}
i \hb \frac{\pa }{\pa t} [ \psi(x_1, t) \phi(x_2, t) ] &=& \bigg[ - \frac{\hb^2}{2m} 
\left( \frac{\pa^2}{\pa x_1^2} + \frac{\pa^2}{\pa x_2^2}\right) + V(x_1) + V(x_2)
\nonumber \\
&+& \frac{1}{2} \frac{\hb \ga}{2i} \left( 
\ln \frac{\psi(x_1, t) }{\psi^*(x_1, t)}
- \left\langle \psi \bigg|  \ln \frac{\psi }{\psi^* } \bigg| \psi \right\rangle
+ \ln \frac{\phi(x_1, t) }{\phi^*(x_1, t)}
- \left\langle \phi \bigg|  \ln \frac{\phi }{\phi^* } \bigg| \phi \right\rangle \right)
\nonumber \\  
&+& \frac{1}{2} \frac{\hb \ga}{2i} \left( 
\ln \frac{\phi(x_2, t) }{\phi^*(x_2, t)}
- \left\langle \phi \bigg|  \ln \frac{\phi }{\phi^* } \bigg| \phi \right\rangle 
+ \ln \frac{\psi(x_2, t) }{\psi^*(x_2, t)}
- \left\langle \psi \bigg|  \ln \frac{\psi }{\psi^* } \bigg| \psi \right\rangle
\right)
  \bigg] 
\nonumber \\  
&\times&  \psi(x_1, t) \phi(x_2, t)    .
\end{eqnarray}
This differential equation then implies that $\psi$ and $\phi$ satisfy two coupled wave equations,
\begin{eqnarray*} 
i \hb \frac{\pa }{\pa t} 
\left\{
\begin{array}{c}
\psi(x, t) \\
\phi(x, t)
\end{array}
\right\}
&=& \bigg[ - \frac{\hb^2}{2m} 
\frac{\pa^2}{\pa x^2} + V(x) 
\nonumber \\
&+& \frac{1}{2} \frac{\hb \ga}{2i} \left( 
\ln \frac{\psi(x, t) }{\psi^*(x, t)}
- \left\langle \psi \bigg|  \ln \frac{\psi }{\psi^* } \bigg| \psi \right\rangle
+ \ln \frac{\phi(x, t) }{\phi^*(x, t)}
- \left\langle \phi \bigg|  \ln \frac{\phi }{\phi^* } \bigg| \phi \right\rangle \right)
\bigg] 
\left\{
\begin{array}{c}
\psi(x, t) \\
\phi(x, t)
\end{array}
\right\}   .
\end{eqnarray*}
However, these equations are not certainly in the form of the SL equation. Thus, the above procedure for symmetrization does not work either.


\newpage


\end{document}